# Attosecond topological interference beyond Floquet–Volkov paths

Hyosub Park[1], J. D. Lee[1]*

[1]Department of Physics and Chemistry, DGIST, Daegu 42988, Republic of Korea
*Corresponding author. Email: jdlee@dgist.ac.kr

**Abstract**
Combining attosecond science with semiconductor physics provides a promising interdisciplinary platform for the attosecond control of semiconducting properties. In each field, the reconstruction of attosecond beating by interference of two-photon transitions (RABBIT) and the time-resolved and angle-resolved photoemission spectroscopy (tr-ARPES) have been central experimental techniques for probing microscopic dynamics of electrons. However, the integration of the two techniques, bridging the two fields, has been missing so far although they have many technical similarities. Here, we develop a theoretical framework that integrates the two techniques based on the second-order time-dependent perturbation theory. Within this framework, we investigate the RABBIT spectroscopy in semiconductors focusing on the recent hallmark of tr-ARPES, namely quantum path interference between Floquet and Volkov states. We identify a novel quantum path incorporating the virtual excitation, which substantially interferes with the Floquet and Volkov states in both the RABBIT spectroscopy and tr-ARPES of semiconductors. Importantly, we find that such virtual excitation channel induces an attosecond topological interference with the Floquet-Volkov paths, encoding topological quantities such as the local Berry curvature into the time shift of RABBIT sidebands, which is directly related to the photoelectron emission delay. Our findings establish a theoretical framework connecting attosecond spectroscopy to tr-ARPES, paving the way for future applications of attosecond science in semiconductors.

**Introduction**

Advancements in attosecond science have enabled real-time tracking of coherent electron motion in atoms, molecules and solids *(1)*. A central experimental technique in attosecond science has been the reconstruction of attosecond beating by interference of two-photon transitions (RABBIT) *(2–4)*. In the standard RABBIT setup, a sequence of attosecond pulses in the XUV regime, called an attosecond pulse train (APT), and an infrared (IR) pulse are used for the pump-probe spectroscopy, and each pulse generates multiple main peaks and sidebands between two main peaks in the photoelectron spectrogram, respectively. In particular, it has been known that the sideband oscillation in the RABBIT spectrogram captures the attosecond-scale evolution of quantum coherence and entanglement of outgoing photoelectrons, which enables us to uncover the intrinsic structure of the system *(5–13)*. A key experimental measure in RABBIT spectroscopy is the attosecond emission delay of the photoelectron, which arises due to the potential scattering and quantum interference of different quantum paths *(14–19)*. Although the RABBIT method has been widely applied to atoms and molecules, there are relatively few investigations of solids. However, it has been found that the electron dynamics in solids is also substantially influenced by periodic potentials *(20–23)*, collective excitations *(24, 25)*, and screening *(26–27)* on the attosecond timescale.

On the other hand, time- and angle-resolved photoemission spectroscopy (tr-ARPES), which is also a pump-probe spectroscopy with a single XUV probe and an IR pump, has provided direct observations of the microscopic dynamics of electrons in semiconductors *(28)*. An important recent achievement in tr-ARPES is the direct observation of Floquet dynamics, which has emerged as a promising nonequilibrium phenomenon, focusing on optical engineering of band structure, Floquet replica states, orbital characters, and topology *(29–34)*. Recently, it has been indicated that tr-ARPES spectra contain not only Floquet states but also Volkov states, which are energetically indistinguishable *(35–42)*. Furthermore, recent experiments have verified that the Floquet state interferes with the Volkov state in graphene *(43,44)*, highlighting that a precise description of all quantum paths is crucial to understand the tr-ARPES data. In this context, applying attosecond techniques such as the RABBIT to semiconducting systems could provide powerful tools for distinguishing microscopic dynamics associated with band structure, pseudospin, topology, etc., possessing a promising potential for attosecond optical control of semiconductors and solid-based qubit operation on an attosecond timescale. To accomplish the connection of RABBIT and tr-ARPES in semiconductors, the precise description of the Floquet and Volkov dynamics in the RABBIT technique is required because these dynamics inevitably occur in general pump-probe interactions in solids due to the absence of a strict optical selection rule. However, there has been a lack of theoretical frameworks that incorporate the tr-ARPES and RABBIT techniques with the precise description of the Floquet-Volkov interference.

In this article, we present a theoretical framework that describes all quantum paths in the tr-ARPES of semiconductors and bridges it with the RABBIT spectroscopy by using the second-order time-dependent perturbation theory. In addition, we analytically identify a novel quantum path involving a virtual conduction state, which overlaps with the Floquet–Volkov spectra in tr-ARPES, demonstrating the incompleteness of the Floquet–Volkov description for tr-ARPES even at the second-order level. Furthermore, we analyze RABBIT sidebands in which quantum interference occurs among the Floquet, Volkov, and virtual excitation paths. We find that this interference induces an attosecond photoemission delay in the RABBIT sidebands, encoding the topology and Berry curvature of the system.

## RESULTS

### Theoretical framework

The overall flow in our theoretical framework is illustrated in Fig. 1. Initially, we develop a comprehensive model covering both the interior and exterior of the material by defining a field operator that includes both bound and unbound states (see section S1 for more details). Note that atomic units are used throughout unless otherwise specified. The total Hamiltonian containing a semiconductor (weakly coupled regime) is written as $H^0 = \sum_{\mathbf{k}} H^0_{\mathbf{k}} = \sum_{\mathbf{k}\mu} E^{\mu}_{\mathbf{k}}\, c^{\dagger}_{\mu\mathbf{k}} c_{\mu\mathbf{k}}$, where $c^{\dagger}_{\mu\mathbf{k}}$ or $c_{\mu\mathbf{k}}$ is the creation and annihilation operator of an electron at momentum $\mathbf{k}$ in $\mu$ band. Here, $\mu$ indicates the valence band ($v$), conduction band ($c$), and unbound ($f$) states, and the Hilbert space of each electron is spanned with the completeness relation $1 = \sum_{\mu\in\{v,c,f\}} |c_{\mu\mathbf{k}}\rangle\langle c_{\mu\mathbf{k}}|$. In this work, the bound states are obtained by the Haldane model (see Methods), which can describe graphene,

2D semiconducting materials (for example, gapped graphene and monolayer TMDs, etc.), effective two-band models of bulk semiconductors with a direct bandgap, and topological materials depending on the breaking of inversion or time-reversal symmetry *(45–48)*. The optical interaction is described by the second-quantized form of incident laser fields, $H_{L\mathbf{k}}(t) = \mathbf{A}(t) \cdot \sum_{\mu\mu'} \mathbf{M}_{\mathbf{k}}^{\mu\mu'} c_{\mu'\mathbf{k}}^{\dagger} c_{\mu\mathbf{k}}$, where $\mathbf{M}_{\mathbf{k}}^{\mu\mu'} = \int d^3r\, \psi_{\mu'\mathbf{k}}^{*}(\mathbf{r})\, \mathbf{k}\, \psi_{\mu\mathbf{k}}(\mathbf{r})$ is the dipole matrix element (see section S1 for more details). The vector potential consists of pump and probe pulses, i.e., $\mathbf{A}(t) = \mathbf{A}^{\mathrm{pump}}(t) + \mathbf{A}^{\mathrm{probe}}(t)$. Each pulse is modelled by a damped cosine function, $\mathbf{A}^{m}(t) = A_0^m \cos[\omega_m(t - \tau_m)] e^{-\eta_m |t-\tau_m|} \vec{\varepsilon}_m$ with $m$ = pump (IR) or probe (XUV), where $\omega_m$, $\eta_m$, and $\tau_m$ are the frequency, inverse-width parameter, and center of the pulse $m$, respectievly. As a convention, $\tau_m$ is set to be zero for $m$ = pump and $\tau_{\mathrm{pump-probe}} \equiv \tau$ for $m$ = probe. Here, $e^{-i\omega_m |t-\tau_m|}$ and $e^{i\omega_m |t-\tau_m|}$ induce absorption and emission processes, respectively.

For the second step of workflow of Fig. 1, we perform the time-dependent perturbation theory considering $H_{\mathbf{k}}^{0}$ as a nonperturbed Hamiltonian with a ground state $|c_{v\mathbf{k}}\rangle$. The time evolution of a quantum state from the ground state is written as $|\psi(t)\rangle = T e^{-i\int dt H_{L\mathbf{k}}^{I}(t)} |c_{v\mathbf{k}}\rangle = \sum_{n=0}^{\infty} |P_n^{\mathbf{k}}(t)\rangle$, where $T$ is the time-ordering operator, with $|P_n^{\mathbf{k}}(t)\rangle = (-i)^n \int dt_1 \cdots dt_n\, H_{L\mathbf{k}}^{I}(t_1) \cdots H_{L\mathbf{k}}^{I}(t_n) |c_{v\mathbf{k}}\rangle$, $H_{L\mathbf{k}}^{I}(t) = e^{iH_{\mathbf{k}}^{0}t} H_{L\mathbf{k}}(t) e^{-iH_{\mathbf{k}}^{0}t}$. In the first-order process, i.e., $|c_{v\mathbf{k}}\rangle \to |c_{\mu\mathbf{k}}\rangle$, all excitations can be coherently treated by inserting the completeness relation,

$$|P_1^{\mathbf{k}}(t)\rangle = \sum_{\mu} \Gamma_{\mathbf{k}\mu}(t) |c_{\mu\mathbf{k}}\rangle, \tag{1}$$

where $\Gamma_{\mathbf{k}\mu}(t) = -i \int_{-\infty}^{t} dt'\, e^{i\Delta E_{\mathbf{k}}^{\mu v} t'} \langle c_{\mu\mathbf{k}} | H_{L\mathbf{k}}(t') | c_{v\mathbf{k}} \rangle$ and $\Delta E_{\mathbf{k}}^{\mu\mu'} = E_{\mathbf{k}}^{\mu} - E_{\mathbf{k}}^{\mu'}$. As the vector potential is given by a superposition of pump and probe pulses, we can rewrite the coefficient of $|P_1^{\mathbf{k}}(t)\rangle$ such as $\Gamma_{\mathbf{k}\mu}(t) = \sum_m \Gamma_{\mathbf{k}\mu}^{m}(t)$ where $\Gamma_{\mathbf{k}\mu}^{m}(t) = -iA_0^m\, \vec{\varepsilon}_m \cdot \mathbf{M}_{\mathbf{k}}^{\mu v} F_{\mathbf{k}}^{\mu v m}(t)$. Here, $F_{\mathbf{k}}^{\mu v m}(t)$ is the first-order transition integral defined by $F_{\mathbf{k}}^{\mu v m}(t) = e^{i\Delta E_{\mathbf{k}}^{\mu v} \tau_m} \left[ \Theta(-t) \frac{e^{i\omega_{\mathbf{k}}^{\mu v m} t + \eta t}}{i\omega_{\mathbf{k}}^{\mu v m} + \eta} + \Theta(t) \left( \frac{2\eta}{\omega_{\mathbf{k}}^{\mu v 2} + \eta^2} + \frac{e^{i\omega_{\mathbf{k}}^{\mu v m} t - \eta t}}{i\omega_{\mathbf{k}}^{\mu v m} - \eta} \right) \right]$ where $\omega_{\mathbf{k}}^{\mu v m} \equiv \Delta E_{\mathbf{k}}^{\mu v} \mp \omega_m$ (see Methods for more details). The sign of $-(+)$ indicates the absorption (emission) process. The second-order process ($| P_2^{\mathbf{k}}(t)\rangle$) is calculated in a similar manner by inserting the completeness relation twice as follows:

$$\left|P_2^{\mathbf{k}}(t)\right\rangle = -\sum_{mm'\mu\mu'} \left(A_0^{m'}\vec{\varepsilon}_{m'} \cdot \mathrm{M}_{\mathrm{k}}^{\mu'\mu}\right)\left(A_0^{m}\vec{\varepsilon}_{m} \cdot \mathrm{M}_{\mathbf{k}}^{\mu v}\right)K_{\mathbf{k}}^{\mu v m \mu' m'}(t)\left|\mathrm{c}_{\mu'\mathbf{k}}\right\rangle, \tag{2}$$

where $K_{\mathbf{k}}^{\mu v m \mu' m'}(t)$ is the time-ordered integral function for the second-order process and details are given in section S2. Each path is generally expressed by $|c_{v\mathbf{k}}\rangle \rightarrow |c_{\mu\mathbf{k}}\rangle \rightarrow |c_{\mu'\mathbf{k}}\rangle$ where $\mu$ and $\mu'$ are indices of intermediate and final states, respectively, and we focus on contributions of $\mathcal{O}\left(A_0^{\mathrm{pump}}A_0^{\mathrm{probe}}\right)$ (see second step in Fig. 1). The summation of $\mu, \mu'$ naturally invokes coherent superposition and interference among the second-order paths in our approach. Spectra of tr-ARPES are obtained by evaluating $\langle P_2^{\mathbf{k}}(\infty)|P_2^{\mathbf{k}}(\infty)\rangle$ that incorporates all possible intermediate states with the final unbound state, i.e., $\mu' = f$, which is assumed to be the free electron state. In principle, there are six second-order paths where the final state is the photoelectron, as shown in fig. S5. Among the six paths, spectra of Floquet and Volkov states are extracted from $|c_{v\mathbf{k}}\rangle \rightarrow |c_{v\mathbf{k}}\rangle \rightarrow |c_{f\mathbf{k}}\rangle$ and $|c_{v\mathbf{k}}\rangle \rightarrow |c_{f\mathbf{k}}\rangle \rightarrow |c_{f\mathbf{k}}\rangle$ paths (see second step Fig. 1), respectively, in accordance with the perturbative treatment adopted in recent tr-ARPES studies *(37–44)*, unlike conventional streaking studies *(5)* under usually much stronger IR field. We note that our approach accurately reproduces recent experiments of Floquet-Volkov interference in graphene *(43, 44)* (see fig. S1).

**Virtual valence-to-conduction transition state**

In the second-order process, there are six quantum pathways including the Floquet and Volkov path, as shown in the upper panel of the second step in Fig. 1. In addition to Floquet and Volkov paths, we indicate another quantum path called virtual valence-to-conduction transition (VVCT) state, which overlaps with Floquet-Volkov features and appreciably contributes to tr-ARPES spectra, denoted by (v) in Fig. 1. This could be better visualized from the first-order time-dependent perturbation theory (i.e., eventually leading to the Fermi golden rule). In Eq. 1, the Floquet state originates from the pump-induced valence-to-valence transition ($|c_{v\mathbf{k}}\rangle \rightarrow |c_{v\mathbf{k}}\rangle$), which is written as $\Gamma_{\mathbf{k}v}^{\mathrm{pump}}(t)|c_{v\mathbf{k}}\rangle = -iA_0^{\mathrm{pump}}\vec{\varepsilon}_{\mathrm{pump}} \cdot \mathbf{M}_{\mathbf{k}}^{vv}F_{\mathbf{k}}^{vv,\mathrm{pump}}(t)|c_{v\mathbf{k}}\rangle \propto e^{\mp i\omega_{\mathrm{pump}}t}|c_{v\mathbf{k}}\rangle$. It

corresponds to the first-order Floquet state of the valence band. According to the time-dependent perturbation theory, the transition integral of the Floquet state due to the probe pulse is proportional to $\delta\left(E_f - E_{\mathbf{k}}^{v} \mp \omega_{\text{pump}} - \omega_{\text{probe}}\right)$, which follows the valence band dispersion shifted by $\omega_{\text{pump}}$ as expected (see Methods). On the other hand, the VVCT state from Eq. 1 is written as $\Gamma_{\mathbf{k}c}^{\text{pump}}(t)|c_{c\mathbf{k}}\rangle = -iA_0^{\text{pump}}\vec{\varepsilon}_{\text{pump}} \cdot \mathbf{M}_{\mathbf{k}}^{cv} F_{\mathbf{k}}^{cv,\text{pump}}(t)|c_{c\mathbf{k}}\rangle \propto e^{i\omega_{\mathbf{k}}^{cv,\text{pump}}t}|c_{c\mathbf{k}}\rangle$ . As shown in Methods, the transition integral for this excitation path contains a delta function, $\delta\left(E_f - E_{\mathbf{k}}^{c} + \omega_{\mathbf{k}}^{cv,\text{pump}} - \omega_{\text{probe}}\right) = \delta\left(E_f - E_{\mathbf{k}}^{v} \mp \omega_{\text{pump}} - \omega_{\text{probe}}\right)$. Notably, the VVCT spectrum also has a valence band dispersion shifted by $\omega_{\text{pump}}$ overlapping with the Floquet and Volkov states in tr-ARPES. The likely engagement of virtual states has been recently addressed *(39, 41)*. Nevertheless, the VVCT state should be strictly distinguished from the Floquet state because its phase modulation is $e^{i\omega_{\mathbf{k}}^{cv,\text{pump}}t}$ not $e^{-i\omega_{\text{pump}}t}$, which is inaccessible within the quasi-static limit where only the Floquet basis is considered *(42)*. The energy dispersion of the VVCT state is directly confirmed by evaluating Eq. 2 only for the path of $|c_{v\mathbf{k}}\rangle \to |c_{c\mathbf{k}}\rangle \to |c_{f\mathbf{k}}\rangle$ in the gapped graphene model *(45)*, as shown in Fig. 2A. The VVCT contribution predominantly occurs at the K point and retains considerable weight even under far-off-resonance conditions, as shown in Fig. 2B (also see figs. S2 and S3), exhibiting a clear virtual feature. In addition, the VVCT state induces the interband transition under the above-resonant excitation (fig. S4). Other second-order pathways (fig. S5) are also included in the total spectra of Fig. 2A but are negligible due to the large energy mismatch.

**RABBIT combined with the tr-ARPES**

We now move our focus to the RABBIT spectroscopy of Floquet-Volkov-VVCT dynamics. Theoretical formulation of RABBIT is realized by considering an attosecond pulse train (APT), which consists of multiple attosecond pulses separated by an interval of neighboring pulses $\Delta T = \pi/\omega_{\text{pump}}$, for the probe (see Fig. 1 and fig. S6). In Fig. 2C, we perform the combined calculation of tr-ARPES and RABBIT by replacing the probe pulse with an APT and obtain characteristic features of the RABBIT spectroscopy, including main peaks and sidebands, as illustrated in the second step in Fig. 1. The main peaks in Fig. 2C are evaluated by $\left\langle P_1^{\mathbf{k}}(\infty)\middle|P_1^{\mathbf{k}}(\infty)\right\rangle$ under the APT probe and it is directly proportional to $1 + \sum_{n=1}^{\infty} e^{-\left|\frac{n\pi}{\omega_{\text{pump}} T_{\text{train}}}\right|} \cos\left(n\pi \frac{\Delta E_{\mathbf{k}}^{fv}}{\omega_{\text{pump}}}\right)$ where $T_{\text{train}}$ is the duration parameter of an APT (see section S4). Here, the summation term directly gives the separated main peaks with an energy interval $2\omega_{\text{pump}}$ (e.g., fig. S7). Meanwhile, the sidebands are calculated by the numerical evaluation of $\langle P_2^{\mathbf{k}}(\infty) | P_2^{\mathbf{k}}(\infty) \rangle$ under an IR pump and APT probe. Corresponding tr-ARPES spectra and the time evolution of tr-ARPES under APT are given in fig. S8 and Movie S1, respectively.

Sidebands at the K point (i.e., $k_x = 0$ in Fig. 2C) are analyzed with respect to the pump-probe delay ($\tau$) in Fig. 3A. In the figure, the intensity of the RABBIT sidebands from Floquet and Volkov states is found to follow the oscillation of $\cos 2\omega_{\text{pump}}\tau$, whereas that of the VVCT state shows a $\pi$ phase shift compared to Floquet and Volkov states, as shown in Fig. 3B. This $\pi$ phase shift comes from the time-integral function $K_{\mathbf{k}}^{\mu v m \mu' m'}(t)$ in the second-order perturbation process (see section S3 and S5). The total RABBIT spectrogram of Fig. 3A shows attosecond-scale temporal shifts of the sideband oscillation, denoted by $\Delta\tau_{\text{K}}$, following $\cos[2\omega_{\text{pump}}(\tau - \Delta\tau_{\text{K}})]$. According to the analog of standard RABBIT spectroscopy *(5)*, $\Delta\tau_{\text{K}}$ is the relative photoelectron emission delay at the K point taking the sideband oscillation of Floquet-Volkov states as a reference. In actual experiment, the reference oscillation would be obtainable by measuring the RABBIT oscillation at a far-off-resonant k-point where the VVCT contribution barely exists. In Fig. 3C, we evaluate the integrated weights of sidebands along the symmetry line of Fig. 2C and find that the VVCT weight is comparable to the Floquet-Volkov weight at the K point. We note that the result is robust to weaker pump strengths at the second-order level. This indicates that the interference among Floquet, Volkov, and VVCT states plays a crucial role in inducing $\Delta\tau_{\text{K}}$, which is experimentally observable by the RABBIT technique.

**Attosecond topological interference in RABBIT sidebands**

In the interference between Floquet-Volkov and VVCT, the VVCT state carries the topological nature of the system through the interband dipole matrix $\vec{\varepsilon}_{\text{pump}} \cdot \mathbf{M}_{\mathbf{k}}^{cv}$, which is directly related to the Berry connection (see Methods). Therefore, we deduce that the topological features affect the Floquet-Volkov-VVCT interference in the RABBIT sidebands. In this context, we construct both topologically trivial and nontrivial systems using the Haldane model (see Methods) with almost the same band structure (see fig. S9). Then, we perform the same calculation as in Fig. 3 in both systems and measure the attosecond-scale time shift in the RABBIT oscillation for every crystal momentum $\mathbf{k}$ (e.g., Fig. 4A). It leads us to obtain the momentum-resolved photoelectron emission delay, $\Delta\tau_{\mathbf{k}}$, for trivial (Figs. 4B and 4C) and nontrivial systems (Figs. 4D and E). In these figures, we observe two noticeable points: (i) time delays are identical for both systems near the K point and (ii) the time delays have opposite signs between the two systems near the K$'$ point. Importantly, it is consistent with the sign change of the Berry curvature in the same region of momentum space, as shown in the inset of Figs. 4B-4E, implying

$$\Delta\tau_{\mathbf{k}}(E_{\mathrm{sb}}) \propto \alpha_{\mathbf{k}}\Omega_{\mathbf{k}}, \tag{3}$$

where $E_{\mathrm{sb}}$ is the energy level of a sideband, $\Omega_{\mathbf{k}}$ the Berry curvature, and $\alpha_{\mathbf{k}}$ a prefactor independent of the topology. Hence, the RABBIT interference between the Floquet-Volkov states and the VVCT state imprints the topological nature of the system on the photoelectron emission delay, which is a characteristic feature of semiconductors that has no direct counterpart in atomic systems. It is consistent with the recent theoretical investigation of the attosecond photoelectron streaking *(49)*.

We analytically examine the interference in RABBIT sidebands in Fig. 4 from Eq. 2. According to the recent study *(49)*, the dipole matrix $\vec{\varepsilon}_{\mathrm{pump}} \cdot \mathbf{M}_{\mathbf{k}}^{cv}$ in the VVCT state has an additional phase factor $e^{i\alpha_{\mathbf{k}}\Omega_{\mathbf{k}}}$ compared to Floquet and Volkov states. As a result, we can simplify the situation of Fig. 4 like $R_{\mathbf{k}} = R_{\mathbf{k}}^{\mathrm{FV}} + R_{\mathbf{k}}^{\mathrm{VVCT}} e^{i\alpha_{\mathbf{k}}\Omega_{\mathbf{k}}}$, where $R_{\mathbf{k}}^{\mathrm{FV}}$ and $R_{\mathbf{k}}^{\mathrm{VVCT}}$ are the scattering amplitudes of the Floquet-Volkov state and the VVCT state, respectively (see section S6 for precise definitions of $R_{\mathbf{k}}^{\mathrm{FV}}$ and $R_{\mathbf{k}}^{\mathrm{VVCT}}$). Then, the RABBIT oscillation is directly calculated by evaluating the sideband intensity, $I_{\mathbf{k}} = \left|R_{\mathbf{k}}^{\mathrm{FV}} + R_{\mathbf{k}}^{\mathrm{VVCT}} e^{i\alpha_{\mathbf{k}}\Omega_{\mathbf{k}}}\right|^2 \equiv I_{\mathbf{k}}^{\mathrm{FV}} + I_{\mathbf{k}}^{\mathrm{VVCT}} + I_{\mathbf{k}}^{\mathrm{FV-VVCT}}$ where $I_{\mathbf{k}}^{\mathrm{FV}} = \left|R_{\mathbf{k}}^{\mathrm{FV}}\right|^2$, $I_{\mathbf{k}}^{\mathrm{VVCT}} = \left|R_{\mathbf{k}}^{\mathrm{VVCT}}\right|^2$, and $I_{\mathbf{k}}^{\mathrm{FV-VVCT}} = 2\mathrm{Re}\left[\left(R_{\mathbf{k}}^{\mathrm{FV}}\right)^* R_{\mathbf{k}}^{\mathrm{VVCT}} e^{i\alpha_{\mathbf{k}}\Omega_{\mathbf{k}}}\right]$. We note that analytic formulas of $I_{\mathbf{k}}^{\mathrm{FV}}$ and $I_{\mathbf{k}}^{\mathrm{VVCT}}$ (see eqs. S26 and S27) well describe the feature of each component RABBIT oscillation in Fig. 3. On the other hand, the pure Floquet-Volkov-VVCT interference contribution, $I_{\mathbf{k}}^{\mathrm{FV-VVCT}}$, is directly proportional to $\cos\omega_{\mathrm{pump}}\tau \cos\Big[2\omega_{\mathrm{pump}}\Big(\tau +$

$\left.\left.\frac{\alpha_{\mathbf{k}}\Omega_{\mathbf{k}}}{2\omega_{\text{pump}}}\right)\right]$ (see eq. S30). In particular, it is found that the factor of $\cos\left[2\omega_{\text{pump}}\left(\tau+\frac{\alpha_{\mathbf{k}}\Omega_{\mathbf{k}}}{2\omega_{\text{pump}}}\right)\right]$ gives rise to the effective shift of the total sideband oscillation depending on $\Omega_{\mathbf{k}}$. As a result, we arrive at the following analytic observations: (i) if Floquet-Volkov is dominant ($\left|R_{\mathbf{k}}^{\text{VVCT}}\right| \ll \left|R_{\mathbf{k}}^{\text{FV}}\right|$), no delay occurs and (ii) if Floquet-Volkov-VVCT interference occurs, a topological time shift of the RABBIT sideband oscillation emerges, in agreement with the numerically measured time shift in Figs. 4B-4E. It implies that the interference between Floquet-Volkov and VVCT states plays a key role in nonvanishing topology-dependent delays of the RABBIT spectroscopy.

## DISCUSSION

We discuss the experimental feasibility of the present RABBIT study. As our formulation is constructed based on the standard RABBIT method and well reproduces recent experimental results *(43,44)*, we expect that the Floquet-Volkov-VVCT interference and its topological nature would be observable in a broad range of RABBIT experiments for 2D materials (graphene, monolayer TMDs, etc.) and bulk semiconductors where the effective two-band model is valid. One of experimentally observable signatures is the relative $\pi$-phase shift between RABBIT oscillations of the Floquet-Volkov path and VVCT path. This measurement would be possible by comparing the RABBIT oscillation at near-resonance and far-off-resonance k-points under an almost resonant pump pulse. On the other hand, the topological phase of the material can be inferred from the momentum-resolved distribution of time delays particularly in the interference regime of the Floquet-Volkov-VVCT states. It is further expected that control of the RABBIT delay would be possible by changing the pump energy, affecting the total interference. It should be noted that our theoretical framework, in principle, has no restriction for the model of semiconductors while this work focuses on the Haldane model for the clear comparison of trivial and topological systems. Because the relation between the interband dipole matrix element and the Berry connection is a general feature, we predict that an analog of the attosecond topological interference induced by the Floquet-Volkov-VVCT states would also emerge in RABBIT sidebands of other semiconductors. We note that the topology-dependent outcomes are robust to higher probe (XUV) energy (see fig. S11), where scattering delays originated from the Coulomb interaction *(52-55)*, Coulomb-laser interaction *(10,54)*, and many-body interactions *(25,55-56)* are negligible. These scattering processes in the low-energy regime could also be accounted for in our theoretical framework by considering more realistic unbound final states, which remains for future work. Nevertheless, the agreement of recent experiments *(43,44)* with our calculation would support the validity of the free-electron approximation in the high energy regime. We also mention that the second-order intensity is gauge invariant for any Bloch gauge transformation, $|v_{\mathbf{k}}/c_{\mathbf{k}}\rangle \rightarrow e^{i\Phi_{v/c}(\mathbf{k})}|v_{\mathbf{k}}/c_{\mathbf{k}}\rangle$ although $\mathbf{M}_{\mathbf{k}}^{cv}$ itself is not gauge-invariant (see section S7).

Next, we remark on the intensity features of VVCT and Floquet states, which play a crucial role in the sideband interference. From Eq. 1, the intensity ratio $R$ between the Floquet state and the VVCT state is estimated by $R = \left|\frac{\Gamma_{\mathbf{k}c}^{\text{pump}}(t)}{\Gamma_{\mathbf{k}v}^{\text{pump}}(t)}\right|^2 = \left|\frac{\vec{\varepsilon}_{\text{pump}}\cdot\mathbf{M}_{\mathbf{k}}^{cv}F_{\mathbf{k}}^{cv,\text{pump}}(t)}{\vec{\varepsilon}_{\text{pump}}\cdot\mathbf{M}_{\mathbf{k}}^{vv}F_{\mathbf{k}}^{vv,\text{pump}}(t)}\right|^2$. In this ratio, $R_i = \left|\left(\vec{\varepsilon}_{\text{pump}}\cdot\mathbf{M}_{\mathbf{k}}^{cv}\right)/\left(\vec{\varepsilon}_{\text{pump}}\cdot\mathbf{M}_{\mathbf{k}}^{vv}\right)\right|^2$ is determined by the optical selection rule, which is related to intrinsic features like pseudospin and surface scattering, while $R_e = \left|F_{\mathbf{k}}^{cv,\text{pump}}(t)/F_{\mathbf{k}}^{vv,\text{pump}}(t)\right|^2$ strongly depends on extrinsic features such as optical conditions of the pump pulse. For $t = -0^+$, we obtain $R_e = \frac{\omega_{\text{pump}}^2+\eta_{\text{pump}}^2}{\omega_{\mathbf{k}}^{cv,\text{pump}^2}+\eta_{\text{pump}}^2}$. If the pulse is extremely short, i.e., $\eta_{\text{pump}} \gg \omega_{\text{pump}}$ (or $\omega_{\mathbf{k}}^{cv,\text{pump}}$), the ratio becomes $R_e \rightarrow 1$. In contrast, if the pulse is long enough (i.e., $\eta_{\text{pump}} \approx 0$), $R_e \rightarrow \left(\omega_{\text{pump}}/\omega_{\mathbf{k}}^{cv,\text{pump}}\right)^2$ in agreement with recent reports *(39, 41)*. In this case, $R_e$ diverges at the resonant transition (VVCT dominant) and converges to zero at the far off-resonant transition (Floquet dominant). It is worth noting that the VVCT path is inhibited in recent graphene experiments *(43,44)* due to the electron doping in the lowest conduction band. The VVCT contribution from the next-lowest conduction band is negligible because the band is placed above 10 eV, which gives $R_e = \mathcal{O}(10^{-2})$. By contrast, previous works which have targeted the Floquet state with a nearly resonant pump energy *(57-60)* may contain the mixture of the Floquet and VVCT state rather than the pure Floquet state. To compare with an atomic system (e.g., a hydrogen atom), the first-order excitation of the $1s$ state is expressed by $\mathbf{A}(t)\cdot\mathbf{p}|\psi_{1s}\rangle = \Gamma_{1s\rightarrow 1s}(t)|\psi_{1s}\rangle + \sum_{nlm}\Gamma_{1s\rightarrow nlm}(t)|\psi_{nlm}\rangle$ with $n \geq 2$. Unlike semiconductors, however, an excitation of the Floquet state corresponding to $\Gamma_{1s\rightarrow 1s}(t)|\psi_{1s}\rangle$ is forbidden due to the strict optical selection rule, whereas the VVCT state $\Gamma_{1s\rightarrow nlm}(t)|\psi_{1s}\rangle$ is suppressed by the large interlevel energy, only the Volkov state being left in the RABBIT sidebands. It proves not only the necessity of Floquet and VVCT states, which are the characteristic features of RABBIT spectroscopy in semiconducting materials, but also the consistency of our approach with the RABBIT spectroscopy in atoms.

In summary, we have developed an analytic theory that incorporates the Floquet-Volkov interference in the tr-ARPES and RABBIT spectroscopy of semiconductors. In agreement with recent observations of the Floquet-Volkov interference, we identify an additional quantum path involving a virtual conduction state (i.e., VVCT), which exhibits a characteristic $\pi$-shift in the RABBIT sidebands and interferes with Floquet-Volkov states in tr-ARPES. Moreover, we show that the interference encodes the topological information in attosecond photoelectron emission delays in the RABBIT sideband structure. These findings establish a theoretical foundation valid for the future attosecond spectroscopic study of semiconducting materials.

## METHODS

### Model Hamiltonian

All systems in the main text are calculated based on the Haldane model, which considers nearest and next-nearest hopping as follows:

$$H_{\mathbf{k}} = -2h_2 \cos\phi \sum_{\boldsymbol{\delta}_2} \cos(\mathbf{k}\cdot\boldsymbol{\delta}_2) + \left[M - 2h_2 \sin\phi \sum_{\boldsymbol{\delta}_2} \sin(\mathbf{k}\cdot\boldsymbol{\delta}_2)\right]\sigma_z$$

$$-h_1 \sum_{\boldsymbol{\delta}_1} \cos(\mathbf{k}\cdot\boldsymbol{\delta}_1)\sigma_x - h_1 \sum_{\boldsymbol{\delta}_1} \sin(\mathbf{k}\cdot\boldsymbol{\delta}_1)\,\sigma_y \tag{4}$$

where $\sigma_i$ $(i = x, y, z)$ is the Pauli matrix and $M$ is the onsite energy of each sublattice basis. Here, $\boldsymbol{\delta}_1$ and $\boldsymbol{\delta}_2$ are the nearest and next-nearest neighbor vectors of graphene, which is fitted with the LDA level ab initio calculation. The topology of the system can be determined by controlling $M$ and the complex phase of the next-nearest hopping parameter $\phi$. This Hamiltonian can illustrate graphene, gapped graphene (topologically trivial), topologically nontrivial systems, and semiconductors with a direct bandgap. For graphene and gapped graphene systems, we set $(h_1, h_2, \phi) = (0.1\ \mathrm{a.u.}, 0, 0)$ and the bandgap is determined by $E_g = 2M$. For the topologically nontrivial system, which has the same band structure as the topologically trivial system, we set $(h_1, M, \phi) = (0.1\ \mathrm{a.u.}, 0, \pi/2)$ and $h_2$ determines the bandgap ( $h_2 = 0.1251\ \mathrm{a.u.} \leftrightarrow E_g = 1.3$ eV and $h_2 = 0.1925\ \mathrm{a.u.} \leftrightarrow E_g = 2.0$ eV). The energy eigenvalues in $H_{\mathbf{k}}^0$ are obtained by diagonalizing the Haldane Hamiltonian, i.e., $H_{\mathbf{k}}^0 = U_{\mathbf{k}} H_{\mathbf{k}} U_{\mathbf{k}}^\dagger = E_{\mathbf{k}}^v |c_{v\mathbf{k}}\rangle\langle c_{v\mathbf{k}}| + E_{\mathbf{k}}^c |c_{c\mathbf{k}}\rangle\langle c_{c\mathbf{k}}|$.

**Transition integral of Floquet and VVCT states**

For a pump pulse, $\mathbf{A}^{\text{pump}}(t) = A_0^{\text{pump}} \cos \omega_{\text{pump}} t\, e^{-\eta|t_{\text{pump}}|} = 2\left(e^{i\omega_{\text{pump}} t - \eta_{\text{pump}}|t|} + e^{-i\omega_{\text{pump}} t - \eta_{\text{pump}}|t|}\right)$, the first-order integral function is defined by

$$F_{\mathbf{k}}^{\mu\nu,\text{pump}}(t) = \int_{-\infty}^{t} d\,t'\, e^{i\omega_{\mathbf{k}}^{\mu\nu,\text{pump}} t' - \eta|t'|}$$

$$= \left[\Theta(-t) \frac{e^{i\omega_{\mathbf{k}}^{\mu\nu,\text{pump}} t + \eta t}}{i\omega_{\mathbf{k}}^{\mu\nu,\text{pump}} + \eta} + \Theta(t) \left( \frac{2\eta}{\omega_{\mathbf{k}}^{\mu\nu,\text{pump}\,2} + \eta^2} + \frac{e^{i\omega_{\mathbf{k}}^{\mu\nu,\text{pump}} t - \eta t}}{i\omega_{\mathbf{k}}^{\mu\nu,\text{pump}} - \eta} \right)\right] \tag{5}$$

where $\omega_{\mathbf{k}}^{\mu\nu,\text{pump}} \equiv \Delta E_{\mathbf{k}}^{\mu\nu} \mp \omega_{\text{pump}}$ and the sign of $-(+)$ indicates the absorption (emission) process. When the pump is long enough (i.e., $\eta_{\text{pump}} \to 0$), it is rewritten as $F_{\mathbf{k}}^{\mu\nu,\text{pump}}(t) \approx \frac{e^{i\omega_{\mathbf{k}}^{\mu\nu,\text{pump}} t}}{i\omega_{\mathbf{k}}^{\mu\nu,\text{pump}}}$. As the Floquet photoelectron state experiences the path of $|c_{v\mathbf{k}}\rangle \to |c_{v\mathbf{k}}\rangle \to |c_{f\mathbf{k}}\rangle$, the first-order state generated by the pump pulse becomes proportional to $e^{\mp i\omega_{\text{pump}} t}|c_{v\mathbf{k}}\rangle$. Let us consider a simplified probe absorption from the valence to the final photoelectron state such as $V = e^{-i\omega_{\text{probe}}}|c_{f\mathbf{k}}\rangle\langle c_{v\mathbf{k}}|$. Then, the transition integral induced by $V$ can be calculated such as

$$\int_{-\infty}^{\infty} d\,t \langle c_{f\mathbf{k}}| e^{iH_{\mathbf{k}}^0 t} V e^{-iH_{\mathbf{k}}^0 t} e^{\mp i\omega_{\text{pump}} t} |c_{v\mathbf{k}}\rangle$$

$$= \int_{-\infty}^{\infty} d\,t\, \langle c_{f\mathbf{k}}| e^{iH_{\mathbf{k}}^0 t} e^{-i\omega_{\text{probe}} t} |c_{f\mathbf{k}}\rangle\langle c_{v\mathbf{k}}| e^{-iH_{\mathbf{k}}^0 t} e^{\mp i\omega_{\text{pump}} t} |c_{v\mathbf{k}}\rangle$$

$$\propto \delta\big(E_f - E_{\mathbf{k}}^{v} \mp \omega_{\text{pump}} - \omega_{\text{probe}}\big). \tag{6}$$

Similarly, the first-order state of the VVCT path ($|c_{v\mathbf{k}}\rangle \rightarrow |c_{c\mathbf{k}}\rangle$) is given as $e^{i(\Delta E_{\mathbf{k}}^{cv} \mp \omega_{\text{pump}})t}|c_{c\mathbf{k}}\rangle$. For this case, we consider the probe absorption from conduction to final photoelectron state such as $V = e^{-i\omega_{\text{probe}}}|c_{f\mathbf{k}}\rangle\langle c_{c\mathbf{k}}|$. Then, the transition integral becomes

$$\int_{-\infty}^{\infty} d\,t\,\langle c_{f\mathbf{k}}|e^{iH_{\mathbf{k}}^{0}t}Ve^{-iH_{\mathbf{k}}^{0}t}e^{i\omega_{\mathbf{k}}^{cv}t}|c_{c\mathbf{k}}\rangle$$

$$\propto \int_{-\infty}^{\infty} d\,t\langle c_{f\mathbf{k}}|\, e^{iH_{\mathbf{k}}^{0}t}e^{-i\omega_{\text{probe}}t}|c_{f\mathbf{k}}\rangle\langle c_{c\mathbf{k}}|e^{-iH_{\mathbf{k}}^{0}t}e^{i(\Delta E_{\mathbf{k}}^{cv} \mp \omega_{\text{pump}})t}|c_{c\mathbf{k}}\rangle$$

$$\propto \delta\big(E_f - E_{\mathbf{k}}^{c} + \Delta E_{\mathbf{k}}^{cv} \mp \omega_{\text{pump}} - \omega_{\text{probe}}\big). \tag{7}$$

As $\Delta E_{\mathbf{k}}^{cv} - E_{\mathbf{k}}^{c} = -E_{\mathbf{k}}^{v}$, we obtain the relation $\int_{-\infty}^{\infty} d\,t\,\langle c_{f\mathbf{k}}|e^{iH_{\mathbf{k}}^{0}t}Ve^{-iH_{\mathbf{k}}^{0}t}e^{i\omega_{\mathbf{k}}^{cv}t}|c_{c\mathbf{k}}\rangle \propto \delta\big(E_f - E_{\mathbf{k}}^{v} \mp \omega_{\text{pump}} - \omega_{\text{probe}}\big)$, which has the same dispersion as the Floquet path.

**Berry connection and VVCT state**

The Berry connection of semiconducting materials is defined in momentum space as $\mathcal{A}_{cv}(\mathbf{k}) = i\langle c_{c\mathbf{k}}|\nabla_{\mathbf{k}}c_{v\mathbf{k}}\rangle$, which is the source of the topological phase. Applying the relation $\nabla_{\mathbf{k}}(H_{\mathbf{k}}|c_{v\mathbf{k}}\rangle) = (\nabla_{\mathbf{k}}H_{\mathbf{k}})|c_{v\mathbf{k}}\rangle + H_{\mathbf{k}}|\nabla_{\mathbf{k}}c_{v\mathbf{k}}\rangle = \nabla_{\mathbf{k}}(E_{\mathbf{k}}^{v}|c_{v\mathbf{k}}\rangle)$, we readily obtain $\boldsymbol{\mathcal{A}}_{cv}(\mathbf{k}) = -i\frac{\langle c_{c\mathbf{k}}|\nabla_{\mathbf{k}}H_{\mathbf{k}}|c_{v\mathbf{k}}\rangle}{E_{\mathbf{k}}^{c} - E_{\mathbf{k}}^{v}}$. Here if we note $\nabla_{\mathbf{k}}H_{\mathbf{k}} = \hat{\mathbf{k}}$, $\langle c_{c\mathbf{k}}|\nabla_{\mathbf{k}}H_{\mathbf{k}}|c_{v\mathbf{k}}\rangle$ is simply proportional to the interband dipole matrix $\vec{\varepsilon}_{\text{probe}} \cdot \mathbf{M}_{\mathbf{k}}^{cv}$ of semiconducting materials, from which the topological information could be directly encoded in the VVCT state through $|c_{v\mathbf{k}}\rangle \rightarrow |c_{c\mathbf{k}}\rangle \rightarrow |c_{f\mathbf{k}}\rangle$. In the Haldane model, it is known that $\langle c_{c\mathbf{k}}|\nabla_{\mathbf{k}}H_{\mathbf{k}}|c_{v\mathbf{k}}\rangle$ has a Berry curvature as a phase factor, i.e., $\langle c_{c\mathbf{k}}|\nabla_{\mathbf{k}}H_{\mathbf{k}}|c_{v\mathbf{k}}\rangle \sim e^{i\alpha_{\mathbf{k}}\Omega_{\mathbf{k}}}$ *(49)*, which directly engages in the sideband oscillation in RABBIT spectrogram.

**Acknowledgments**

**Funding:** This work was supported by the Basic Science Research Program (RS-2025-00553164) and the EDISON2.0 program (RS-2023-00253716) through the National Research Foundation of Korea (NRF), funded by the Ministry of Science and ICT.

**Author contributions:** H.P. and J.D.L. designed the study. H.P. did analytic and numerical calculations. H.P. and J.D.L. discussed results and wrote the paper.

**Competing interests:** All authors declare they have no competing interests.

**Data, code, and materials availability:** All data and code needed to evaluate and reproduce the results in the paper are present in the paper and/or the Supplementary Materials. This study did not generate any new materials.

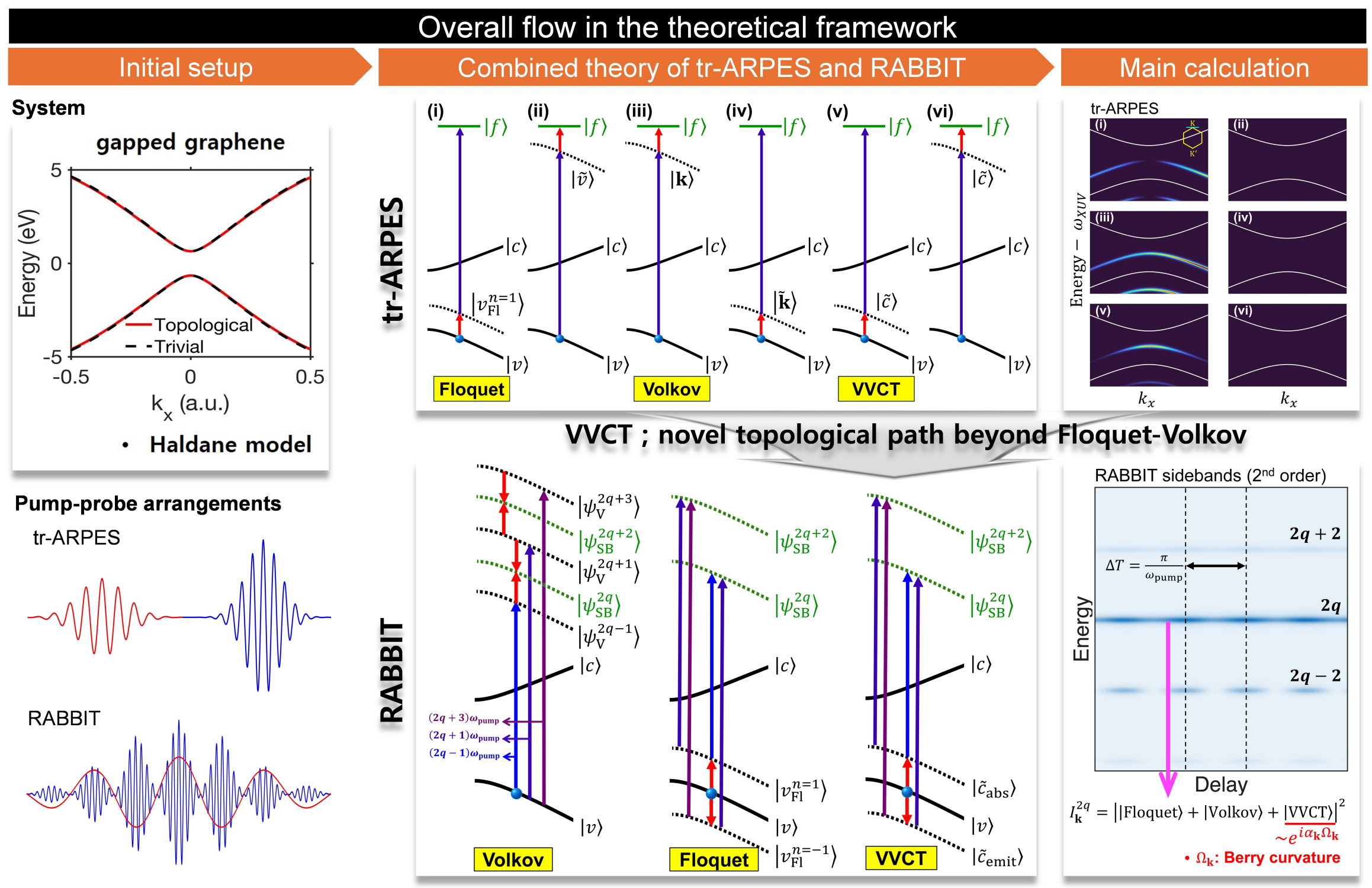

**Fig. 1. Overall flow in theoretical framework.** The workflow consists of initial setup, developing combined theory of tr-ARPES and RABBIT, and main calculations. The initial setup constructs a system including an unperturbed Hamiltonian (such as gapped graphene and Haldane model) and light-matter interaction with pump (red) and probe (blue) pulses. The probe pulse is a single (multiple) attosecond pulse(s) for tr-ARPES (RABBIT). In the combined theory, the tr-ARPES includes not only Floquet ($|v\rangle \to |v_{\mathrm{Fl}}^{n=\pm1}\rangle \to |f\rangle$) and Volkov ($|v\rangle \to |\mathbf{k}\rangle \to |f\rangle$) paths but also virtual valence-to-conduction transition (VVCT, ($|v\rangle \to |\tilde{c}\rangle \to |f\rangle$)) path and extended to the RABBIT spectroscopy, which gives rise to sidebands ($|\psi_{SB}^{2q}\rangle$) inside main peaks (e.g., $|\psi_{V}^{2q-1}\rangle$), where $|\mathbf{k}\rangle$ is the free-electron state and $q$ is an integer. In Floquet and VVCT paths, we omit the odd-order photoelectron states, which are identical to those of Volkov path, for better visualization. Red and blue arrows indicate pump (IR) and probe (XUV) interactions, respectively. In main calculations, spectra of tr-ARPES and sideband oscillation of RABBIT are directly evaluated, which contains the interference of Floquet-Volkov-VVCT paths. In particular, the VVCT-induced topological feature, which contains the Berry curvature as a phase *(49)*, is examined in the interference of RABBIT sidebands by both numerical and analytical perspectives.

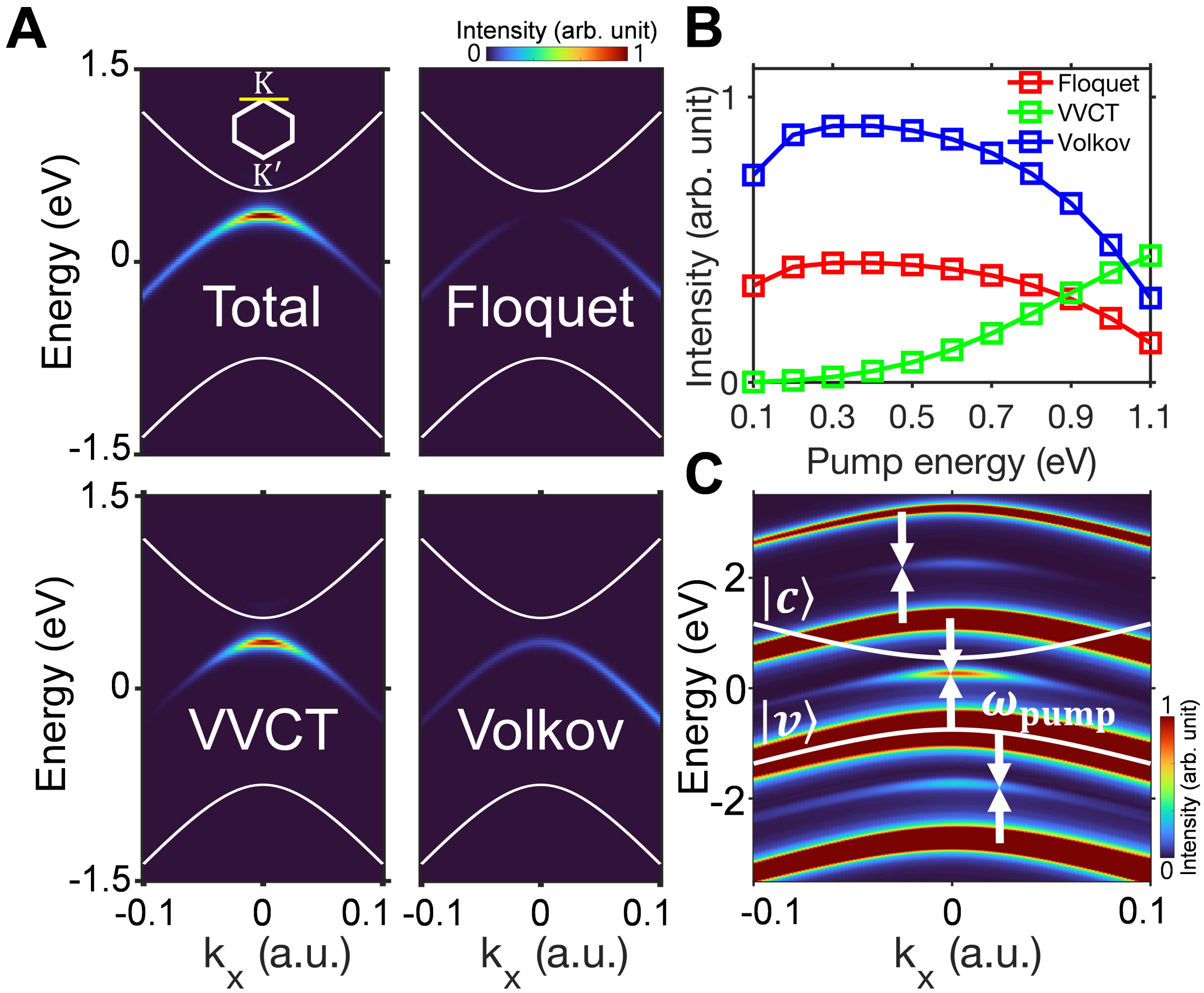

**Fig. 2. tr-ARPES evaluated by the analytic formula.** (**A**) tr-ARPES at $\tau_{\mathrm{pump-probe}} = 0$ fs along the yellow line. Total, Floquet, Volkov, and VVCT spectra are calculated with the x-polarized pump ($\omega_{\mathrm{pump}} = 1.0$ eV and $\eta_{\mathrm{pump}}^{-1} = 36.1$ fs) and probe ($\omega_{\mathrm{probe}} = 100$ eV and $\eta_{\mathrm{probe}}^{-1} = 21.6$ fs) pulses for the gapped graphene model with a bandgap of 1.3 eV. We adopt $A_0^{\mathrm{probe}} = 10^{-6}$ a.u. and $A_0^{\mathrm{pump}} = 5.3 \times 10^{-3}$ a.u. (i.e., a field strength of 1 MV/cm) and set the polar angle of both pulses to be 8.3° *(44)*. (**B**) Integrated weights of Floquet, Volkov, and VVCT, which are normalized by the total weight with respect to the pump energy. (**C**) tr-ARPES in the RABBIT implementation at $\tau_{\mathrm{pump-probe}} = 0$ fs. The pump pulse is same as in (**A**) and the APT probe is given by a sum of 29 damped cosine functions, where the time interval $\Delta T$ is $\pi/\omega_{\mathrm{pump}}$ (see fig. S6). $T_{\mathrm{train}}$ is set to be 7 fs.

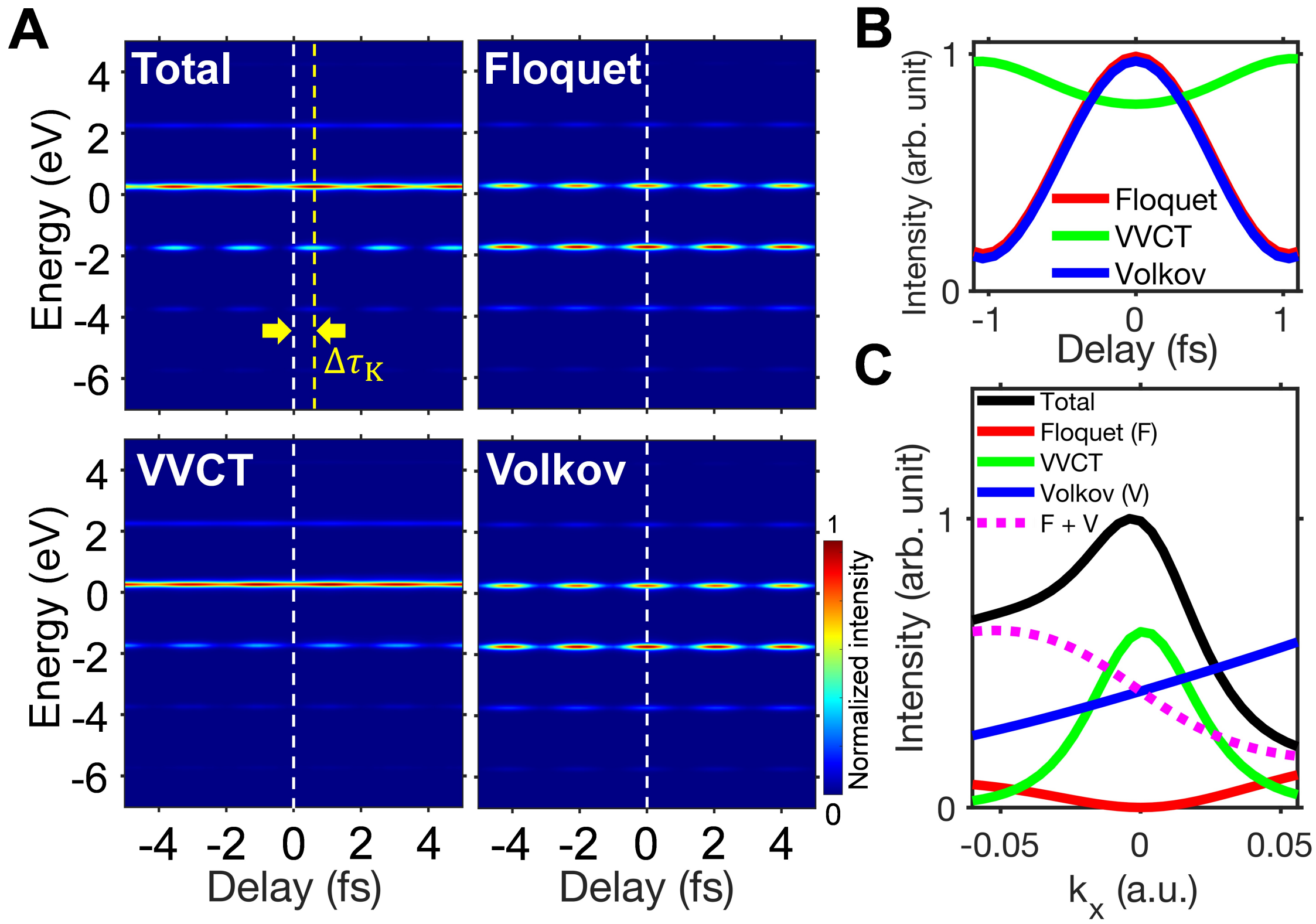

**Fig. 3. RABBIT spectrogram containing the Floquet-Volkov-VVCT interference.** (**A**) Total RABBIT spectrograms and separate contributions from Floquet, Volkov, and VVCT states at the K point with respect to the pump-probe delay $\tau_{\text{pump}-\text{probe}}$. The intensities are normalized by the maximum of each spectrogram. The vertical white dashed lines indicate $\tau_{\text{pump-probe}} = 0$ fs. (**B**) RABBIT oscillations from Floquet, Volkov, and VVCT states around 0 eV of (**A**). (**C**) Momentum-resolved and energy-integrated weights of sidebands in Fig. 2C.

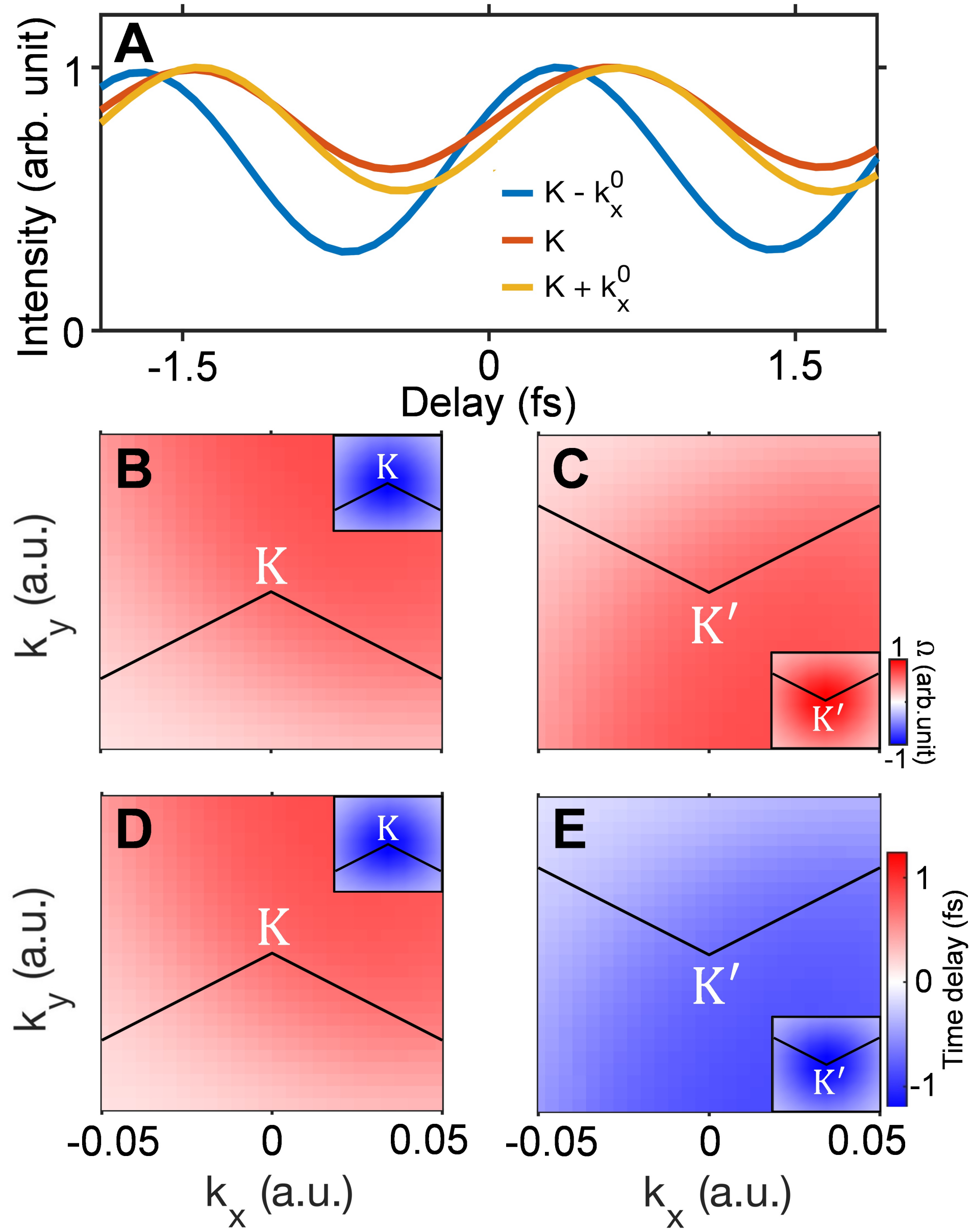

**Fig. 4. Time shifts of RABBIT sideband oscillation and their topological features.** (**A**) Oscillation of the intensity in the RABBIT sideband at the points $\mathrm{K}-(k_x^0, 0)$, K, and $\mathrm{K}+(k_x^0, 0)$ ($k_x^0 = 0.04$ a.u.). (**B**)-(**E**) Photoelectron emission delays around the point of trivial (**B**, **C**) and nontrivial (**D**, **E**) systems are displayed around K and K′ points. Signs of the corresponding Berry curvatures are given in the insets. $\alpha_{\mathbf{k}} < 0$ and $\alpha_{\mathbf{k}} > 0$ are understood around the K and K′ points, respectively.

## Supplementary Materials for

# Attosecond topological interference beyond Floquet–Volkov paths

Hyosub Park and J. D. Lee*

*Corresponding author. Email: jdlee@dgist.ac.kr

## S1. Construction of the second-quantized Hamiltonian

A many-electron Hamiltonian of a solid system can be expressed by

$$H = \sum_i \frac{p_i^2}{2m_e} + \sum_i \frac{P_i^2}{2m_p} - \sum_{i,j} \frac{e^2}{|\mathbf{R}_i - \mathbf{r}_j|} + \frac{1}{2}\sum_{i,j} \frac{e^2}{|\mathbf{r}_i - \mathbf{r}_j|} + \frac{1}{2}\sum_{i,j} \frac{e^2}{|\mathbf{R}_i - \mathbf{R}_j|} \tag{S1}$$

where $m_e$ mass of an electron, $m_p$ mass of an atomic core, $\mathbf{R}_i$ position of an atomic core, and $\boldsymbol{r}_i$ position of an electron. Since $m_p$ is much larger than $m_e$, $\mathbf{R}_i$ and $P_i$ are usually assumed to be fixed values. We adopt atomic units throughout the whole text and assume that the second and last terms in the above equation are a reference energy that can be omitted. Moreover, the third term reduces to a one-body interaction consisting of the background crystal potential for the electron. We define the one-body part of the Hamiltonian as

$$H^0 = \sum_i \frac{p_i^2}{2} + U. \tag{S2}$$

where $U = \sum_i U_i(\mathbf{r}_i)$, $U_i(\mathbf{r}_i) = -\sum_j \frac{e^2}{|\mathbf{R}_j - \mathbf{r}_i|}$ and firth them in the Eq. (S1), which consists of two-body electron-electron interaction, is denoted by $V = \frac{1}{2}\sum_{i,j} \frac{e^2}{|\mathbf{r}_i - \mathbf{r}_j|}$. Then, the total Hamiltonian is reduced to $H = H^0 + V$. For condensed matter systems, $U$ characterizes the crystal structure of the system and $V$ determines the correlation properties of the material, such as screening. Here, we assume that the material is a semiconductor or band insulator, enabling us to ignore $V$. Then, the Hamiltonian of each single electron (index $i$) is defined by $H_i^0 = \frac{p_i^2}{2} + U_i(\mathbf{r}_i)$. Note that $H_i$ is identical for all $i$ and independent of each other, hence solving the solution of a single provides a solution of the total system. Therefore, we restrict the problem into a single electron Hamiltonian and change denotations such as $H_i \rightarrow H$ and $\mathbf{r}_i, \mathbf{p}_i \rightarrow \mathbf{r}, \mathbf{p}$.

Typically, it is useful to deal with the Hamiltonian in the second-quantized form. Let $\{\psi_\lambda(\mathbf{r})\}$ be the set of eigenfunctions of $H^0$. Then, the second quantized Hamiltonian for $\{\psi_\lambda(\mathbf{r})\}$ is written as

$$H^0 = \sum_\lambda E_\lambda \, \hat{c}_\lambda^\dagger \hat{c}_\lambda \,, \tag{S3}$$

where $E_\lambda = \int d^3 r \psi_\lambda^\dagger(\mathbf{r}) H_0 \psi_\lambda(\mathbf{r})$ is the eigenvalue of $H_0$ and $\hat{c}_\lambda$ is the corresponding ladder operator. Let $\lambda = (\mathbf{k}, \mu)$ where $\mathbf{k}$ and $\mu$ are electron momentum and band index, respectively When the two-body interaction is negligible, an eigenfunction obeys Bloch's theorem due to the spatial periodicity of $U$ and thus the basis wavefunction satisfies $\psi_{\mu\mathbf{k}}(\mathbf{r} + \mathbf{R}) = e^{i\mathbf{k}\cdot\mathbf{R}} \psi_{\mu\mathbf{k}}(\mathbf{r})$ where $\mathbf{R}$ is the lattice vector. Note that the set of $\{\psi_{\mu\mathbf{k}}(\mathbf{r})\}$ contains both bound and unbound states. The bound states are countable, describing standing waves inside the potential, while the unbound states are continuous as they are formed due to the scattering between unbounded electrons and the potential. Both bound and unbound states are, in principle, possible to be obtained by solving the eigenvalue problem of the time-independent Schrödinger equation. Then, the field operator can be separately expressed by

$$\hat{\Psi}(\mathbf{r}) = \sum_{\mu \leq w, \mathbf{k}} \hat{c}_{\mu\mathbf{k}} \, \psi_{\mu\mathbf{k}}(\boldsymbol{r}) + \sum_{\mu > w, \mathbf{k}} \hat{c}_{\mu\mathbf{k}} \, \psi_{\mu\mathbf{k}}(\boldsymbol{r}). \tag{S4}$$

Here, $\mu$ is the index for energy bands including bound states and unbound states, respectively, and $w$ is a boundary index satisfying $E_{w\mathbf{k}} > W$, $E_{w\mathbf{k}} \leq W$ where $W$ is the work function. Utilizing this field operator, the second quantization of the total Hamiltonian is evaluated as

$$H^0 = \sum_{\mu \leq w,\mathbf{k}} E_{\mathbf{k}}^{\mu}\, \hat{c}_{\lambda\mathbf{k}}^{\dagger} \hat{c}_{\lambda\mathbf{k}} + \sum_{\mu > w,\mathbf{k}} E_{\mathbf{k}}^{\mu}\, \hat{c}_{\lambda\mathbf{k}}^{\dagger} \hat{c}_{\lambda\mathbf{k}}, \quad \text{(S5)}$$

where each electron, which has a different momentum, is dealt with independently because the two-body potential is neglected. As a result, it is possible to focus on each single-electron Hamiltonian separately below

$$H_{\mathbf{k}}^0 = \sum_{\mu \leq w} E_{\mathbf{k}}^{\mu}\, \hat{c}_{\lambda\mathbf{k}}^{\dagger} \hat{c}_{\lambda\mathbf{k}} + \sum_{\mu > w} E_{\mathbf{k}}^{\mu}\, \hat{c}_{\lambda\mathbf{k}}^{\dagger} \hat{c}_{\lambda\mathbf{k}}. \quad \text{(S6)}$$

The bound terms are usually described by the tight-binding model. On the other hand, the unbound states are usually approximated as free-electron or time-reversed LEED states because it is difficult to obtain an exact final state. Here, we adopt the free-electron approximation.

The optical interaction can be described by the minimal coupling, $H(t) = \sum_{\mathbf{k}}\left[\frac{\left(\mathbf{k}-q\mathbf{A}(t)\right)^2}{2} + U\right] \approx \sum_{\mathbf{k}}\left[\frac{k^2}{2} + U\right] - q\sum_{\mathbf{k}} \hat{\mathbf{k}} \cdot \mathbf{A}(t)$ where $q = -1$. The latter term in the right-hand side equation is the light-matter interaction, $H_L(t)$, and its second quantization is

$$\begin{aligned} H_L(t) &= \mathbf{A}(t) \cdot \sum_{\mathbf{k}\mu\mu'} \mathbf{M}_{\mathbf{k}}^{\mu\mu'} c_{\mu'\mathbf{k}}^{\dagger} c_{\mu\mathbf{k}} \\ &\equiv \sum_{\mathbf{k}} H_{L\mathbf{k}}(t), \end{aligned} \quad \text{(S7)}$$

where $H_{L\mathbf{k}}(t) = \mathbf{A}(t) \cdot \sum_{\mu\mu'} \mathbf{M}_{\mathbf{k}}^{\mu\mu'} c_{\mu'\mathbf{k}}^{\dagger} c_{\mu\mathbf{k}}$ and $\mathbf{M}_{\mathbf{k}}^{\mu\mu'} = \int d^3r\, \psi_{\mu'\mathbf{k}}^{*}(\mathbf{r})\, \hat{\mathbf{k}} \psi_{\mu\mathbf{k}}(\mathbf{r})$. The light-matter interaction of the bound-to-bound transition can be described by the Peierls substitution of a tight-binding Hamiltonian, i.e., $H_{\mathbf{k}}^0 \to H_{\mathbf{k}-\mathbf{A}(t)}^0$ in the sublattice basis. Specifically, we use the Haldane model,

$$H_{\mathbf{k}}^{0} = \begin{bmatrix} M - 2h_2 \sum_{\boldsymbol{\delta}_2} \cos(\mathbf{k}\cdot\boldsymbol{\delta}_2 - \phi) & -h_1 \sum_{\boldsymbol{\delta}_1} e^{-i\mathbf{k}\cdot\boldsymbol{\delta}_1} \\ -h_1 \sum_{\boldsymbol{\delta}_1} e^{i\mathbf{k}\cdot\boldsymbol{\delta}_1} & -M + 2h_2 \sum_{\boldsymbol{\delta}_2} \cos(\mathbf{k}\cdot\boldsymbol{\delta}_2 - \phi) \end{bmatrix}, \tag{S8}$$

which can describe graphene, gapped graphene (topologically trivial), and topologically nontrivial systems. Here, $\boldsymbol{\delta}_1$ and $\boldsymbol{\delta}_2$ are the nearest and next-nearest vectors of graphene. For graphene and gapped graphene systems, we set $(h_1, h_2, \phi) = (0.1\ \text{a.u.},\ 0, 0)$ and the bandgap is determined by $E_g = 2M$. For the topologically nontrivial system, we set $(h_1, M, \phi) = (0.1\ \text{a.u.}, 0, \pi/2)$ and $h_2$ determines the bandgap ($h_2 = 0.1251\ \text{a.u.} \leftrightarrow E_g = 1.3$ eV and $h_2 = 0.1925\ \text{a.u.} \leftrightarrow E_g = 2.0$ eV). The dipole matrix, $\mathbf{M}_{\mathbf{k}}^{\mu\mu'}$, is then obtained from the first-order Taylor expansion of the Hamiltonian for the vector potential. On the other hand, matrix elements of bound-to-unbound and unbound-to-unbound transitions are obtained under the free-electron approximation for the unbound state, i.e., $\mathbf{M}_{\mathbf{k}}^{\mu\mu'} = \mathbf{k} \int d^3r\, \psi_{\mu'\mathbf{k}}^{*}(\mathbf{r})\, \psi_{\mu\mathbf{k}}(\mathbf{r})$.

## S2. Second-order time-dependent perturbation theory with multiple pulses

The multiple input pulses are modeled by the sum of damped cosine functions, $\mathbf{A}(t) = \sum_m A_0^m \cos\omega_m t_m\, e^{-\eta_m |t_m|} \vec{\varepsilon}_m = \sum_m \frac{A_0^m}{2} \left(e^{i\omega_m t_m - \eta_m |t_m|} + e^{-i\omega_m t_m - \eta_m |t_m|}\right) \vec{\varepsilon}_m$, where $t_m = t - \tau_m$. Here, $m$ is the index of pulses and $\tau_m$ is the central time of the $m$-th pulse. For convenience, let us change the denotation of the field strength, such as $\frac{A_0^m}{2} \rightarrow A_0^m$. As described in the main text, the first-order order time-dependent perturbation theory (TDPT) with the multiple pulses is then written as $\left|P_1^{\mathbf{k}}(t)\right\rangle = \sum_\mu \Gamma_{\mathbf{k}\mu}(t) \left|c_{\mu\mathbf{k}}\right\rangle$ where $\Gamma_{\mathbf{k}\mu}(t) = -i \sum_m A_0^m\, \vec{\varepsilon}_m \cdot \mathbf{M}_{\mathbf{k}}^{\mu v} F_{\mathbf{k}}^{\mu v m}(t) \equiv \sum_m \Gamma_{\mathbf{k}\mu}^{m}$, including the pulse index, i.e., $F_{\mathbf{k}}^{\mu v}(t) \rightarrow F_{\mathbf{k}}^{\mu v m}(t)$. Here, we redefine the first-order time integral function such as $F_{\mathbf{k}}^{\mu v m}(t) \equiv e^{i\Delta E_{\mathbf{k}}^{\mu v} \tau_m} \bar{F}_{\mathbf{k}}^{\mu v m}(t)$ where $\bar{F}_{\mathbf{k}}^{\mu v m}(t) = \int_{-\infty}^{t} dt' e^{i\omega_{\mathbf{k}}^{\mu v m} t'_m - \eta_m |t'_m|} =$ $\Theta(-t_m)\frac{e^{i\omega_{\mathbf{k}}^{\mu v m}t_m+\eta_m t_m}}{i\omega_{\mathbf{k}}^{\mu v m}+\eta_m}+\Theta(t_m)\left(\frac{2\eta_m}{\omega_{\mathbf{k}}^{\mu v m 2}+\eta_m^2}+\frac{e^{i\omega_{\mathbf{k}}^{\mu v m}t_m-\eta_m t_m}}{i\omega_{\mathbf{k}}^{\mu v m}-\eta_m}\right)$, $\Delta E_{\mathbf{k}}^{\lambda\lambda'}\equiv E_{\mathbf{k}}^{\lambda}-E_{\mathbf{k}}^{\lambda'}$, and $\omega_{\mathbf{k}}^{\lambda\lambda' m}\equiv \Delta E_{\mathbf{k}}^{\lambda\lambda'}-\omega_m$. Note that the emission process is obtained by substituting $\omega_m$ to $-\omega_m$. Then, the second-order term can be straightforwardly written as

$$\left|P_2^{\mathbf{k}}(t)\right\rangle=-\sum_{m,\mu\in\{v,c\}}A_0^m\vec{\varepsilon}_m\cdot\mathbf{M}_{\mathbf{k}}^{\mu v}\int_{-\infty}^{t}dt'\,F_{\mathbf{k}}^{\mu v m}(t')\,e^{iH_{\mathbf{k}}^0 t'}H_{L\mathbf{k}}(t')e^{-iH_{\mathbf{k}}^0 t'}\left|c_{\mu\mathbf{k}}\right\rangle. \quad \text{(S9)}$$

Here, the integrand can be expanded by inserting the identity operator defined in the main text,

$$\begin{aligned} &e^{iH_{\mathbf{k}}^0 t'}H_{L\mathbf{k}}(t')e^{-iH_{\mathbf{k}}^0 t'}\left|c_{\mu\mathbf{k}}\right\rangle \\ &\quad=e^{i(E_{\mathbf{k}}^{v}-E_{\mathbf{k}}^{\mu})}\langle c_{v\mathbf{k}}|H_{L\mathbf{k}}(t')\left|c_{\mu\mathbf{k}}\right\rangle|c_{v\mathbf{k}}\rangle+\sum_{\mu'}e^{i\left(E_{\mathbf{k}}^{\mu'}-E_{\mathbf{k}}^{\mu}\right)t'}\left\langle c_{\mu'\mathbf{k}}\right|H_{L\mathbf{k}}(t')\left|c_{\mu\mathbf{k}}\right\rangle\left|c_{\mu'\mathbf{k}}\right\rangle. \end{aligned} \quad \text{(S10)}$$

Each former and latter term corresponds to the recombination and scattering of the excited electron, respectively, and we only focus on the scattering process now. Then, we obtain

$$\begin{aligned} \left|P_2^{\mathbf{k}}(t)\right\rangle&=-\sum_{m\mu}\vec{\varepsilon}_m\cdot\mathbf{M}_{\mathbf{k}}^{\mu v} \\ &\quad\times\sum_{\mu'}\int_{-\infty}^{t}dt'F_{\mathbf{k}}^{\mu v m}(t')\left\langle c_{\mu'\mathbf{k}}\right|e^{iH_{\mathbf{k}}^0 t'}H_{L\mathbf{k}}(t')e^{-iH_{\mathbf{k}}^0 t'}\left|c_{\mu\mathbf{k}}\right\rangle\left|c_{\mu'\mathbf{k}}\right\rangle. \end{aligned} \quad \text{(S11)}$$

This second-order process can be divided into four channels,

$$\begin{aligned} &\text{(i) } |c_{v\mathbf{k}}\rangle\to|c_{v\mathbf{k}}\rangle\to|c_{v\mathbf{k}}\rangle \\ &=-\sum_m A_0^m\vec{\varepsilon}_m\cdot\mathbf{M}_{\mathbf{k}}^{vv}\int_{-\infty}^{t}dt'F_{\mathbf{k}}^{vvm}(t')(\langle c_{v\mathbf{k}}|H_{L\mathbf{k}}(t')|v_{\mathbf{k}}\rangle)\,|c_{v\mathbf{k}}\rangle \\ &=-\sum_{mm'}(A_0^m\vec{\varepsilon}_m\cdot\mathbf{M}_{\mathbf{k}}^{vv})\left(A_0^{m'}\vec{\varepsilon}_{m'}\cdot\mathbf{M}_{\mathbf{k}}^{vv}\right)\int_{-\infty}^{t}dt'F_{\mathbf{k}}^{vvm}(t')\,e^{-i\omega_{m'}t'-\eta_{m'}\left|t'_{m'}\right|}|c_{v\mathbf{k}}\rangle \\ &\text{(ii) } |c_{v\mathbf{k}}\rangle\to\left|c_{\mu\mathbf{k}}\right\rangle\to|c_{v\mathbf{k}}\rangle\text{ where }\mu\neq v \\ &=-\sum_{m\mu}\left(A_0^m\vec{\varepsilon}_m\cdot\mathbf{M}_{\mathbf{k}}^{\mu v}\right)\int_{-\infty}^{t}dt'F_{\mathbf{k}}^{\mu v m}(t')\left(\langle c_{v\mathbf{k}}|e^{iH_{\mathbf{k}}^0 t'}H_{L\mathbf{k}}(t')e^{-iH_{\mathbf{k}}^0 t'}\left|c_{\mu\mathbf{k}}\right\rangle\right)|c_{v\mathbf{k}}\rangle \\ &=-\sum_{mm'}\left(A_0^m\vec{\varepsilon}_m\cdot\mathbf{M}_{\mathbf{k}}^{\mu v}\right)\left(A_0^{m'}\vec{\varepsilon}_{m'}\cdot M_{\mathbf{k}}^{v\mu}\right)\int_{-\infty}^{t}dt'F_{\mathbf{k}}^{\mu v m}(t')e^{i\omega_{\mathbf{k}}^{v\mu m'}t'-\eta_{m'}\left|t'_{m'}\right|}\,|c_{v\mathbf{k}}\rangle \\ &\text{(iii) } |c_{v\mathbf{k}}\rangle\to|c_{v\mathbf{k}}\rangle\to\left|c_{\mu\mathbf{k}}\right\rangle\text{ where }\mu\neq v \end{aligned} \quad \text{(S12)}$$

$$= -\sum_m \left(A_0^m \vec{\varepsilon}_m \cdot \mathbf{M}_{\mathbf{k}}^{vv}\right) \sum_{n'\mu} \left(A_0^{m'} \vec{\varepsilon}_{m'} \cdot M_{\mathbf{k}}^{\mu v}\right) \int_{-\infty}^{t} dt' F_{\mathbf{k}}^{vvm}(t') e^{i\omega_{\mathbf{k}}^{\mu v m'} t' - \eta_{m'} \left|t'_{m'}\right|} |c_{v\mathbf{k}}\rangle$$

(iv) $|c_{v\mathbf{k}}\rangle \rightarrow |c_{\mu\mathbf{k}}\rangle \rightarrow |c_{\mu'\mathbf{k}'}\rangle$ where $\mu \neq v$ & $\mu' \neq v$

$$= -\sum_{\substack{mm' \\ \mu\mu'}} \left(A_0^m \vec{\varepsilon}_m \cdot \mathbf{M}_{\mathbf{k}}^{\mu v}\right) \left(A_0^{m'} \vec{\varepsilon}_{m'} \cdot M_{\mathbf{k}}^{\mu'\mu}\right) \int_{-\infty}^{t} dt' F_{\mathbf{k}}^{\mu v m}(t') e^{i\omega_{\mathbf{k}}^{\mu'\mu m'} t' - \eta_{m'} \left|t'_{m'}\right|} |c_{\mu'\mathbf{k}}\rangle \ .$$

The photoemitted state is then obtained by taking the final state to be an unbound state (i.e., $\mu'$: unbound), such as following:

$$\left|P_2^{\mathbf{k}}(t)\right\rangle = -\sum_{mm'\mu\mu'} \left(A_0^{m'} \vec{\varepsilon}_{m'} \cdot \mathbf{M}_{\mathbf{k}}^{\mu'\mu}\right) \left(A_0^m \vec{\varepsilon}_m \cdot \mathbf{M}_{\mathbf{k}}^{\mu v}\right) K_{\mathbf{k}}^{\mu v m \mu' m'}(t) \left|c_{\mu'\mathbf{k}}\right\rangle, \qquad \text{(S13)}$$

where $K_{\mathbf{k}}^{\mu v m \mu' m'}(t) \equiv \int_{-\infty}^{t} dt' F_{\mathbf{k}}^{\mu v m}(t') e^{i\omega_{\mathbf{k}}^{\mu'\mu m'} t' - \eta_{m'} \left|t'_{m'}\right|}$.

## S3. Time integral function in the two-photon process

The time integral in the second-order perturbation process is defined as follows (see section S2):

$$K_{\mathbf{k}}^{\mu v m \mu' m'}(t) = e^{i\Delta E_{\mathbf{k}}^{\mu'\mu} \tau_{m'}} e^{i\Delta E_{\mathbf{k}}^{\mu v} \tau_m} \overline{K}_{\mathbf{k}}^{\mu v m \mu' m'}(t), \qquad \text{(S14)}$$

where

$$\begin{aligned} \overline{K}_{\mathbf{k}}^{\mu v m \mu' m'}(t) &= \int_{-\infty}^{t} dt' \Theta(-t'_m) \frac{e^{i\omega_{\mathbf{k}}^{\mu v m} t'_m + \eta_m t'_m}}{i\omega_{\mathbf{k}}^{\mu v m} + \eta_m} e^{i\omega_{\mathbf{k}}^{\mu'\mu m'} t'_{m'} - \eta_{m'} \left|t'_{m'}\right|} \\ &+ \int_{-\infty}^{t} dt' \, \Theta(t'_m) \left( \frac{2\eta_m}{\omega_{\mathbf{k}}^{\mu v m 2} + \eta_m^2} + \frac{e^{i\omega_{\mathbf{k}}^{\mu v m} t'_m - \eta_m t'_m}}{i\omega_{\mathbf{k}}^{\mu v m} - \eta_m} \right) e^{i\omega_{\mathbf{k}}^{\mu'\mu m'} t'_{m'} - \eta_{m'} \left|t'_{m'}\right|}. \end{aligned} \qquad \text{(S15)}$$

It should be calculated separately depending on the sign of $t'_m = t' - \tau_m$ and $t'_{m'} = t' - \tau_{m'}$.

Case 1: If $t > \tau_m$, both first and second terms in the integrand survive. The step functions break the integration rage such as below:

$$\overline{K}_{\mathbf{k}}^{\mu v m \mu' m'}(t) = \int_{-\infty}^{\tau_m} dt' \frac{e^{i\omega_{\mathbf{k}}^{\mu v m} t'_m + \eta_m t'_m}}{i\omega_{\mathbf{k}}^{\mu v m} + \eta_m} e^{i\omega_{\mathbf{k}}^{\mu'\mu m'} t'_{m'} - \eta_{m'} \left|t'_{m'}\right|} \qquad \text{(S16)}$$

$$+\int_{\tau_m}^{t} dt' \left(\frac{2\eta_m}{\omega_{\mathbf{k}}^{\mu\nu m^2}+\eta_m^2}+\frac{e^{i\omega_{\mathbf{k}}^{\mu\nu m}t'_m-\eta_m t'_m}}{i\omega_{\mathbf{k}}^{\mu\nu m}-\eta_m}\right) e^{i\omega_{\mathbf{k}}^{\mu'\mu m'}t'_{m'}-\eta_{m'}\left|t'_{m'}\right|} .$$

Note that $\omega_{\mathbf{k}}^{\lambda'\lambda m}+\omega_{\mathbf{k}}^{\lambda''\lambda' m'}=\omega_{\mathbf{k}}^{\lambda''\lambda m}-\omega_{m'}$. This integral is then divided into the following three cases in accordance with the sign of $t'_{m'}$.

(i) If $\tau_m \geq \tau_{m'}$

$$\bar{K}_{\mathbf{k}}^{\mu\nu m\mu' m'}(t)=\int_{-\infty}^{\tau_{m'}} dt' \frac{e^{i\omega_{\mathbf{k}}^{\mu\nu m}t'_m+\eta_m t'_m}}{i\omega_{\mathbf{k}}^{\mu\nu m}+\eta_m} e^{i\omega_{\mathbf{k}}^{\mu'\mu m'}t'_{m'}+\eta_{m'}t'_{m'}}$$

$$+\int_{\tau_{m'}}^{\tau_m} dt' \frac{e^{i\omega_{\mathbf{k}}^{\mu\nu m}t'_m+\eta_m t'_m}}{i\omega_{\mathbf{k}}^{\mu\nu m}+\eta_m} e^{i\omega_{\mathbf{k}}^{\mu'\mu m'}t'_{m'}-\eta_{m'}t'_{m'}}$$

$$+\int_{\tau_m}^{t} dt' \left(\frac{2\eta_m}{\omega_{\mathbf{k}}^{\mu\nu m^2}+\eta_m^2}+\frac{e^{i\omega_{\mathbf{k}}^{\mu\nu m}t'_m-\eta_m t'_m}}{i\omega_{\mathbf{k}}^{\mu\nu m}-\eta_m}\right) e^{i\omega_{\mathbf{k}}^{\mu'\mu m'}t'_{m'}-\eta_{m'}t'_{m'}}$$

$$=\frac{1}{i\omega_{\mathbf{k}}^{\mu\nu m}+\eta_m}\frac{e^{\left(i\omega_{\mathbf{k}}^{\mu\nu m}+\eta_m\right)\left(\tau_{m'}-\tau_m\right)}}{i\left(\omega_{\mathbf{k}}^{\mu\nu m}+\omega_{\mathbf{k}}^{\mu'\mu m'}\right)+\eta_m+\eta_{m'}}+\frac{1}{i\omega_{\mathbf{k}}^{\mu\nu m}+\eta_m}\frac{e^{\left(i\omega_{\mathbf{k}}^{\mu'\mu m'}-\eta_{m'}\right)\left(\tau_m-\tau_{m'}\right)}-e^{\left(i\omega_{\mathbf{k}}^{\mu\nu m}+\eta_m\right)\left(\tau_{m'}-\tau_m\right)}}{i\left(\omega_{\mathbf{k}}^{\mu\nu m}+\omega_{\mathbf{k}}^{\mu'\mu m'}\right)+\eta_m-\eta_{m'}}$$

$$+\frac{2\eta_m}{\omega_{\mathbf{k}}^{\mu\nu m^2}+\eta_m^2}\frac{e^{\left(i\omega_{\mathbf{k}}^{\mu'\mu m'}-\eta_{m'}\right)t_{m'}}-e^{\left(i\omega_{\mathbf{k}}^{\mu'\mu m'}-\eta_{m'}\right)\left(\tau_m-\tau_{m'}\right)}}{i\omega_{\mathbf{k}}^{\mu'\mu m'}-\eta_{m'}}$$

$$+\frac{1}{i\omega_{\mathbf{k}}^{\mu\nu m}-\eta_m}\frac{e^{\left(i\omega_{\mathbf{k}}^{\mu\nu m}-\eta_m\right)t_m+\left(i\omega_{\mathbf{k}}^{\mu'\mu m'}-\eta_{m'}\right)t_{m'}}-e^{\left(i\omega_{\mathbf{k}}^{\mu'\mu m'}-\eta_{m'}\right)\left(\tau_m-\tau_{m'}\right)}}{i\left(\omega_{\mathbf{k}}^{\mu\nu m}+\omega_{\mathbf{k}}^{\mu'\mu m'}\right)-\eta_m-\eta_{m'}}$$

(ii) If $\tau_m < \tau_{m'} < t$

$$\bar{K}_{\mathbf{k}}^{\mu\nu m\mu' m'}(t)=\int_{-\infty}^{\tau_m} dt' \frac{e^{i\omega_{\mathbf{k}}^{\mu\nu m}t'_m+\eta_m t'_m}}{i\omega_{\mathbf{k}}^{\mu\nu m}+\eta_m} e^{i\omega_{\mathbf{k}}^{\mu'\mu m}t'_{m'}+\eta_{m'}t'_{m'}}$$

$$+\int_{\tau_m}^{\tau_{m'}} dt' \left(\frac{2\eta_m}{\omega_{\mathbf{k}}^{\mu\nu m^2}+\eta_m^2}+\frac{e^{i\omega_{\mathbf{k}}^{\mu\nu m}t'_m-\eta_m t'_m}}{i\omega_{\mathbf{k}}^{\mu\nu m}-\eta_m}\right) e^{i\omega_{\mathbf{k}}^{\mu'\mu m'}t'_{m'}+\eta_{m'}t'_{m'}}$$

$$+\int_{\tau_{m'}}^{t} dt' \left(\frac{2\eta_m}{\omega_{\mathbf{k}}^{\mu\nu m^2}+\eta_m^2}+\frac{e^{i\omega_{\mathbf{k}}^{\mu\nu m}t'_m-\eta_m t'_m}}{i\omega_{\mathbf{k}}^{\mu\nu m}-\eta_m}\right) e^{i\omega_{\mathbf{k}}^{\mu'\mu m'}t'_{m'}-\eta_{m'}t'_{m'}}$$

$$= \frac{1}{i\omega_{\mathbf{k}}^{\mu v m}+\eta_m} \frac{e^{\left(i\omega_{\mathbf{k}}^{\mu'\mu m'}+\eta_{m'}\right)(\tau_m-\tau_{m'})}}{i\left(\omega^{\mathbf{k}}_{\mu v m}+\omega_{\mathbf{k}}^{\mu'\mu m'}\right)+\eta_m+\eta_{m'}} + \frac{2\eta_m}{\omega_{\mathbf{k}}^{\mu v m^2}+\eta_m^2} \frac{1-e^{\left(i\omega_{\mathbf{k}}^{\mu'\mu m'}+\eta_{m'}\right)(\tau_m-\tau_{m'})}}{i\omega_{\mathbf{k}}^{\square\square'\mu m'}+\eta_{m'}}$$

$$+ \frac{1}{i\omega_{\mathbf{k}}^{\mu v m}-\eta_m} \frac{e^{\left(i\omega_{\mathbf{k}}^{\mu v m}-\eta_m\right)(\tau_{m'}-\tau_m)} - e^{\left(i\omega_{\mathbf{k}}^{\mu'\mu m'}+\eta_{m'}\right)(\tau_m-\tau_{m'})}}{i\left(\omega_{\mathbf{k}}^{\mu v m}+\omega_{\mathbf{k}}^{\mu'\mu m'}\right)-\eta_m+\eta_{m'}}$$

$$+ \frac{2\eta_m}{\omega_{\mathbf{k}}^{\mu v m^2}+\eta_m^2} \frac{e^{\left(i\omega_{\mathbf{k}}^{\mu'\mu m'}-\eta_{m'}\right)(t-\tau_{m'})}-1}{i\omega_{\mathbf{k}}^{\mu'\mu m'}-\eta_{m'}}$$

$$+ \frac{1}{i\omega_{\mathbf{k}}^{\mu v m}-\eta_m} \frac{e^{\left(i\omega_{\mathbf{k}}^{\mu v m}-\eta_m\right)t_m+\left(i\omega_{\mathbf{k}}^{\mu'\mu m'}-\eta_{m'}\right)t_{m'}} - e^{\left(i\omega_{\mathbf{k}}^{\mu v m}-\eta_m\right)(\tau_{m'}-\tau_m)}}{i\left(\omega_{\mathbf{k}}^{\mu v m}+\omega_{\mathbf{k}}^{\mu'\mu m'}\right)-\eta_m-\eta_{m'}}$$

(iii) If $\tau_m < t < \tau_{m'}$

$$\overline{K}_{\mathbf{k}}^{\mu v m\mu' m'}(t) = \int_{-\infty}^{\tau_m} dt' \frac{e^{i\omega_{\mathbf{k}}^{\mu v m}t'_m+\eta_m t'_m}}{i\omega_{\mathbf{k}}^{\mu v m}+\eta_m} e^{i\omega_{\mathbf{k}}^{\mu'\mu m'}t'_{m'}+\eta_{m'}t'_{m'}}$$

$$+ \int_{\tau_m}^{t} dt' \left( \frac{2\eta_m}{\omega_{\mathbf{k}}^{\mu v m^2}+\eta_m^2} + \frac{e^{i\omega_{\mathbf{k}}^{\mu v m}t'_m-\eta_m t'_m}}{i\omega_{\mathbf{k}}^{\mu v m}-\eta_m} \right) e^{i\omega_{\mathbf{k}}^{\mu'\mu m'}t'_{m'}+\eta_{m'}t'_{m'}}$$

Case 2: If $t < \tau_m$, only the first term in the integrand survives.

$$\overline{K}_{\mathbf{k}}^{\mu v m\mu' m'}(t) = \int_{-\infty}^{t} dt' \frac{e^{i\omega_{\mathbf{k}}^{\mu v m}t'_m+\eta_m t'_m}}{i\omega_{\mathbf{k}}^{\mu v m}+\eta_m} e^{i\omega_{\mathbf{k}}^{\mu'\mu m'}t'_{m'}-\eta_{m'}\left|t'_{m'}\right|} \quad . \qquad \text{(S17)}$$

(i) If $\tau_m < \tau_{m'}$ or $t < \tau_{m'} < \tau_m$

$$\overline{K}_{\mathbf{k}}^{\mu v m\mu' m'}(t) = \int_{-\infty}^{t} dt' \frac{e^{i\omega_{\mathbf{k}}^{\mu v m}t'_m+\eta_m t'_m}}{i\omega_{\mathbf{k}}^{\mu v m}+\eta_m} e^{i\omega_{\mathbf{k}}^{\mu'\mu m'}t'_{m'}+\eta_{m'}t'_{m'}}$$

$$= \frac{1}{i\omega_{\mathbf{k}}^{\mu v m}+\eta_m} \frac{e^{i\left(\omega_{\mathbf{k}}^{\mu v m}t_m+\omega_{\mathbf{k}}^{\mu'\mu m'}t_{m'}\right)+\eta_m t'_m+\eta_{m'}t'_{m'}}}{i\left(\omega_{\mathbf{k}}^{\mu v m}+\omega_{\mathbf{k}}^{\mu'\mu m'}\right)+\eta_m+\eta_{m'}}$$

(i) If $\tau_{m'} < t < \tau_m$

$$\overline{K}_{\mathbf{k}}^{\mu v m\mu' m'}(t) = \int_{-\infty}^{\tau_{m'}} dt' \frac{e^{i\omega_{\mathbf{k}}^{\mu v m}t'_m+\eta_m t'_m}}{i\omega_{\mathbf{k}}^{\mu v m}+\eta_m} e^{i\omega_{\mathbf{k}}^{\mu'\mu m'}t'_{m'}+\eta_{m'}t'_{m'}}$$

$$+\int_{\tau_{n'}}^{t} dt' \frac{e^{i\omega_{\mathbf{k}}^{\mu\nu m} t'_m + \eta_m t'_m}}{i\omega_{\mathbf{k}}^{\mu\nu m} + \eta_m} e^{i\omega_{\mathbf{k}}^{\mu'\mu m'} t'_{m'} - \eta_{m'} t'_{m'}}.$$

For a special case, $t \to \infty$, the above integrations are reduced to the following two cases:

(i) If $\tau_m \geq \tau_{m'}$

$$\bar{K}_{\mathbf{k}}^{\mu\nu m\mu' m'}(\infty) = \frac{1}{i\omega_{\mathbf{k}}^{\mu\nu m} + \eta_m} \frac{e^{\left(i\omega_{\mathbf{k}}^{\mu\nu m} + \eta_m\right)\left(\tau_{m'} - \tau_m\right)}}{i\left(\omega_{\mathbf{k}}^{\mu\nu m} + \omega_{\mathbf{k}}^{\mu'\mu m'}\right) + \eta_m + \eta_{m'}}$$

$$+\frac{1}{i\omega_{\mathbf{k}}^{\mu\nu m} + \eta_m} \frac{e^{\left(i\omega_{\mathbf{k}}^{\mu'\mu m'} - \eta_{m'}\right)\left(\tau_m - \tau_{m'}\right)} - e^{\left(i\omega_{\mathbf{k}}^{\mu\nu m} + \eta_m\right)\left(\tau_{m'} - \tau_m\right)}}{i\left(\omega_{\mathbf{k}}^{\mu\nu m} + \omega_{\mathbf{k}}^{\mu'\mu m'}\right) + \eta_m - \eta_{m'}}$$

$$-\frac{2\eta_m}{\omega_{\mathbf{k}}^{\mu\nu m 2} + \eta_m^2} \frac{e^{\left(i\omega_{\mathbf{k}}^{\mu'\mu m'} - \eta_{m'}\right)\left(\tau_m - \tau_{m'}\right)}}{i\omega_{\mathbf{k}}^{\mu'\mu m'} - \eta_{m'}} - \frac{1}{i\omega_{\mathbf{k}}^{\mu\nu m} - \eta_m} \frac{e^{\left(i\omega_{\mathbf{k}}^{\mu'\mu m'} - \eta_{m'}\right)\left(\tau_m - \tau_{m'}\right)}}{i\left(\omega_{\mathbf{k}}^{\mu\nu m} + \omega_{\mathbf{k}}^{\mu'\mu m'}\right) - \eta_m - \eta_{m'}}$$

(ii) If $\tau_m < \tau_{m'}$

$$\bar{K}_{\mathbf{k}}^{\mu\nu m\mu' m'}(\infty) = \frac{1}{i\omega_{\mathbf{k}}^{\mu\nu m} + \eta_m} \frac{e^{\left(i\omega_{\mathbf{k}}^{\mu'\mu m'} + \eta_{m'}\right)\left(\tau_m - \tau_{m'}\right)}}{i\left(\omega_{\mathbf{k}}^{\mu\nu m} + \omega_{\mathbf{k}}^{\mu'\mu m'}\right) + \eta_m + \eta_{m'}} + \frac{2\eta_m}{\omega_{\mathbf{k}}^{\mu\nu m 2} + \eta_m^2} \frac{1 - e^{\left(i\omega_{\mathbf{k}}^{\mu'\mu m'} + \eta_{m'}\right)\left(\tau_m - \tau_{m'}\right)}}{i\omega_{\mathbf{k}}^{\mu'\mu m'} + \eta_{m'}}$$

$$+\frac{1}{i\omega_{\mathbf{k}}^{\mu\nu m} - \eta_m} \frac{e^{\left(i\omega_{\mathbf{k}}^{\mu\nu m} - \eta_m\right)\left(\tau_{m'} - \tau_m\right)} - e^{\left(i\omega_{\mathbf{k}}^{\mu'\mu m'} + \eta_{m'}\right)\left(\tau_m - \tau_{m'}\right)}}{i\left(\omega_{\mathbf{k}}^{\mu\nu m} + \omega_{\mathbf{k}}^{\mu'\mu m'}\right) - \eta_m + \eta_{m'}} - \frac{2\eta_m}{\omega_{\mathbf{k}}^{\mu\nu m 2} + \eta_m^2} \frac{1}{i\omega_{\mathbf{k}}^{\mu'\mu m'} - \eta_{m'}}$$

$$-\frac{1}{i\omega_{\mathbf{k}}^{\mu\nu m} - \eta_m} \frac{e^{\left(i\omega_{\mathbf{k}}^{\mu\nu m} - \eta_m\right)\left(\tau_{m'} - \tau_m\right)}}{i\left(\omega_{\mathbf{k}}^{\mu\nu m} + \omega_{\mathbf{k}}^{\mu'\mu m'}\right) - \eta_m - \eta_{m'}}.$$

## S4. Main peaks in RABBIT generated by an attosecond pulse train (APT)

Main peaks in RABBIT come from the first-order excitation induced by an APT probe. This APT is the sequential injection of multiple attosecond pulses, and it can be modeled by considering multiple damped cosine functions in our approach. Let $m$ be an integer index for the multiple damped cosine pulses. Then, an APT can be realized by $\mathbf{A}_{\mathrm{APT}}(t) = \sum_m A_0^m \cos \omega_m t_m \, e^{-\eta_m |t_m|} \vec{\varepsilon}_m$ where $t_m \equiv t - \tau_m$ and $m = 1, 2, \dots, N_{\text{pulse}}$ ($N_{\text{pulse}}$: the number of cosine pulses belonging to APT). Following the standard RABBIT method, $\omega_m$ ($= \omega_{\text{probe}}$), $\eta_m$, and $\vec{\varepsilon}_m$ are the same for every $m$. Instead, each damped cosine pulse is timely separated by $\tau_m = \tau_0 + m\Delta T$ where $\tau_0$ is the center of the APT and $\Delta T = \frac{\pi}{\omega_{\text{pump}}}$, which makes each cosine pulse placed at different extrema of the pump amplitude. In the pump-probe scheme, the center of the APT is set to be $\tau_0 = \tau_{\text{pump-probe}}$. In this section, however, we set that $\tau_0 = 0$ for simplicity. In actual experiments, the whole APT also has a damped factor, and it is modelled by assuming $A_0^m = A_0 e^{-\left|\frac{m\Delta T}{T_{\text{train}}}\right|}$, where $T_{\text{train}}$ is a duration parameter of the whole APT. The first-order time-integral function for the APT at $t \to \infty$ is then written by $F_{\mathbf{k}}^{\mu v m}(\infty) = \frac{2\eta_{\text{probe}}}{\omega_{\mathbf{k}}^{\mu v 2} + \eta_{\text{probe}}^2} e^{im\Delta E_{\mathbf{k}}^{\mu v}\Delta T}$ where $\omega_{\mathbf{k}}^{\mu v} = \Delta E_{\mathbf{k}}^{\mu\mu'} - \omega_{\text{probe}}$. Since the APT is the probe pulse, the excited state is the photoelectron state (i.e., $\mu = f$). Then, the first-order contribution under the APT, $\Gamma_{\mathbf{k}f}(\infty) = -i\vec{\varepsilon}_{\text{probe}} \cdot \mathbf{M}_{\mathbf{k}}^{fv} \sum_m A_0^m F_{\mathbf{k}}^{fvm}(\infty)$, is written as

$$\Gamma_{\mathbf{k}f}(\infty) = \frac{-2i\eta\chi A_0 \vec{\varepsilon}_{\text{probe}} \cdot \mathbf{M}_{\mathbf{k}}^{fv}}{\omega_{\mathbf{k}}^{\mu v 2} + \eta_{\text{probe}}^2} \sum_m e^{-\left|\frac{m\Delta T}{T_{\text{train}}}\right|} e^{im\Delta E_{\mathbf{k}}^{fv}\Delta T}$$
$$\propto 1 + 2\sum_{n=1} e^{-\left|\frac{n\pi}{\omega_{\text{pump}} T_{\text{train}}}\right|} \cos\left(n\pi \frac{\Delta E_{\mathbf{k}}^{fv}}{\omega_{\text{pump}}}\right). \quad \text{(S18)}$$

Main peaks of the RABBIT method come from the summation of $\cos\left(n\pi \frac{\Delta E_{\mathbf{k}}^{fv}}{\omega_{\text{pump}}}\right)$. Mathematically, $\sum_{n=1} \cos\left(n\pi \frac{x}{x_0}\right)$ gives multiple peaks separated by $2x_0$ (e.g., see Fig. S7), explaining the main peaks of RABBIT calculations in the main text. Note that the damped factor, $e^{-\left|\frac{n\pi}{\omega_{\text{pump}} T_{\text{train}}}\right|}$, suppresses small wiggling between main peaks of $\sum_{n=1} \cos\left(n\pi \frac{\Delta E_{\mathbf{k}}^{fv}}{\omega_{\text{pump}}}\right)$. A much sharper energy separation between neighboring main peaks is also available when we use a Gaussian envelope for the damping factor of an APT, i.e., $A_0^m = A_0 e^{-\left(\frac{m\Delta T}{T_{\mathrm{train}}}\right)^2}$, while such change barely affects results in the main text.

## S5. Analysis of RAABIT sidebands

The second-order integral function for $t \to \infty$ a is obtained by calculating the following formula (see section S3),

$$
\begin{aligned}
\bar{K}_{\mathbf{k}}^{\mu v m \mu' m'}(\infty) = & \frac{1}{i\omega_{\mathbf{k}}^{\mu v m}+\eta_m} \frac{e^{\left(i\omega_{\mathbf{k}}^{\mu v m}+\eta_m\right)\left(\tau_{m'}-\tau_m\right)}}{i\left(\omega_{\mathbf{k}}^{\mu v m}+\omega_{\mathbf{k}}^{\mu' \mu m'}\right)+\eta_m+\eta_{m'}} \\
& + \frac{1}{i\omega_{\mathbf{k}}^{\mu v m}+\eta_m} \frac{e^{\left(i\omega_{\mathbf{k}}^{\mu' \mu m'}-\eta_{m'}\right)\left(\tau_m-\tau_{m'}\right)} - e^{\left(i\omega_{\mathbf{k}}^{\mu v m}+\eta_m\right)\left(\tau_{m'}-\tau_m\right)}}{i\left(\omega_{\mathbf{k}}^{\mu v m}+\omega_{\mathbf{k}}^{\mu' \mu m'}\right)+\eta_m-\eta_{m'}} \\
& - \frac{2\eta_m}{\omega_{\mathbf{k}}^{\mu v m 2}+\eta_m^2} \frac{e^{\left(i\omega^{\mathbf{k}}_{\mu' \mu m'}-\eta_{m'}\right)\left(\tau_m-\tau_{m'}\right)}}{i\omega_{\mathbf{k}}^{\mu' \mu m'}-\eta_{m'}} - \frac{1}{i\omega_{\mathbf{k}}^{\mu v m}-\eta_m} \frac{e^{\left(i\omega_{\mathbf{k}}^{\mu' \mu m'}-\eta_{m'}\right)\left(\tau_m-\tau_{m'}\right)}}{i\left(\omega_{\mathbf{k}}^{\mu v m}+\omega_{\mathbf{k}}^{\mu' \mu m'}\right)-\eta_m-\eta_{m'}}.
\end{aligned}
\tag{S19}
$$

Let $\Delta\tau_{mm'} \equiv \tau_m - \tau_{m'} = \pm\tau_{\mathrm{pump-probe}}$. At sidebands, say $E_f = E_{\mathbf{k}}^v + \omega_{\mathrm{probe}} + (2k-1)\omega_{\mathrm{pump}}$ ($k$: integer), there are absorption and emission processes induced by the pump pulse. In each case, we have a relation, $\omega_{\mathbf{k}}^{\mu v m} + \omega_{\mathbf{k}}^{f \mu m'} = E_f - E_{\mathbf{k}}^v - \omega_{\mathrm{probe}} \mp \omega_{\mathrm{pump}} = (2k-1)\omega_{\mathrm{pump}} \mp \omega_{\mathrm{pump}}$. Let us only consider the case where $k = 1$, without loss of generality. In this case, we obtain $\omega_{\mathbf{k}}^{\mu v m} + \omega_{\mathbf{k}}^{f \mu m'} = 2(l-1)\omega_{\mathrm{pump}}$ where $l = 1$ and $l = 2$ for absorption ($-\omega_{\mathrm{pump}}$) and emission ($+\omega_{\mathrm{pump}}$) processes, respectively. Then,

$$\bar{K}_{\mathbf{k}}^{\mu\nu mfm'}(\infty) = e^{-i\omega_{\mathbf{k}}^{\mu\nu m}\Delta\tau_{mm'}}\left[\frac{1}{i\omega_{\mathbf{k}}^{\mu\nu m}+\eta_m}\frac{e^{-\eta_m\Delta\tau_{mm'}}}{2i(l-1)\omega_{\text{pump}}+\eta_m+\eta_{m'}}\right.$$
$$+\frac{1}{i\omega_{\mathbf{k}}^{\mu\nu m}+\eta_m}\frac{e^{(2i(l-1)\omega_{\text{pump}}-\eta_{m'})\Delta\tau_{mm'}}-e^{-\eta_m\Delta\tau_{mm'}}}{2i(l-1)\omega_{\text{pump}}+\eta_m-\eta_{m'}}$$
$$-\frac{2\eta_m}{\omega_{\mathbf{k}}^{\mu\nu m2}+\eta_m^2}\frac{e^{(2i(l-1)\omega_{\text{pump}}-\eta_{m'})\Delta\tau_{mm'}}}{i\left(2(l-1)\omega_{\text{pump}}-\omega_{\mathbf{k}}^{\mu\nu m}\right)-\eta_{m'}} \tag{S20}$$
$$\left.-\frac{1}{i\omega_{\mathbf{k}}^{\mu\nu m}-\eta_m}\frac{e^{(i2i(l-1)\omega_{\text{pump}}-\eta_{m'})\Delta\tau_{mm'}}}{2i(l-1)\omega_{\text{pump}}-\eta_m-\eta_{m'}}\right]$$

**Case 1.** For absorption process ($l = 1$), Eq. (S19) is written as

$$\bar{K}_{\mathbf{k}}^{\mu\nu mfm'}(\infty) = e^{-i\omega_{\mathbf{k}}^{\mu\nu m}\Delta\tau_{mm'}}\left[\frac{1}{i\omega_{\mathbf{k}}^{\mu\nu m}+\eta_m}\frac{e^{-\eta_m\Delta\tau_{mm'}}}{\eta_m+\eta_{m'}}\right.$$
$$+\frac{1}{i\omega_{\mathbf{k}}^{\mu\nu m}+\eta_m}\frac{e^{-\eta_{m'}\Delta\tau_{mm'}}-e^{-\eta_m\Delta\tau_{mm'}}}{\eta_m-\eta_{m'}} \tag{S21}$$
$$\left.+\frac{2\eta_m}{\omega_{\mathbf{k}}^{\mu\nu m2}+\eta_m^2}\frac{e^{-\eta_{m'}\Delta\tau_{mm'}}}{i\omega_{\mathbf{k}}^{\mu\nu m}+\eta_{m'}}+\frac{1}{i\omega_{\mathbf{k}}^{\mu\nu m}-\eta_m}\frac{e^{-\eta_{m'}\Delta\tau_{mm'}}}{\eta_m+\eta_{m'}}\right]$$

For absorption processes of Floquet, Volkov, and VVCT, the following energy relations are obtained:

Floquet: $\omega_{\mathbf{k}}^{\mu\nu m} = \omega_{\mathbf{k}}^{\nu\nu,\text{pump}} = -\omega_{\text{pump}}$

Volkov: $\omega_{\mathbf{k}}^{\mu\nu m} = -\omega_{\mathbf{k}}^{\mu'\mu m'} = -\omega_{\mathbf{k}}^{ff,\text{pump}} = \omega_{\text{pump}}$

VVCT: $\omega_{\mathbf{k}}^{\mu\nu m} = \omega_{\mathbf{k}}^{c\nu,\text{pump}} = \Delta E_{\mathbf{k}}^{c\nu} - \omega_{\text{pump}}$.

Note that $\omega_{\text{pump}} = 1.0$ eV, $\eta_{\text{pump}} \approx \mathcal{O}(0.1\text{ eV})$ and $\eta_{\text{probe}} \approx \mathcal{O}(10\text{ eV})$ for the pulse conditions used in the main text. Then, we approximate $\eta_{\text{pump}} \approx 0$ and $\frac{\omega_{\text{pump}}}{\eta_{XUV}} \approx 0$ such as

- Floquet & VVCT ($m = \text{pump}, m' = \text{probe}, \eta_m \to 0$):

$$K_{\mathbf{k}}^{\mu v m \mu' m'}(\infty) \approx -\frac{2ie^{-i\omega_{\mathbf{k}}^{\mu v m}\Delta\tau_{mm'}}}{\omega_{\mathbf{k}}^{\mu v m}\eta_{m'}}$$

- Volkov ($m = \text{probe}$, $m' = \text{pump}$, $\eta_{m'} \to 0$):

$$K_{\mathbf{k}}^{\mu v m \mu' m'}(\infty) \approx -\frac{2ie^{-i\omega_{\mathbf{k}}^{\mu v m}\Delta\tau_{mm'}}}{\omega_{\mathbf{k}}^{\mu v m}\eta_{m}}.$$

Therefore, the integral function is also proportional to

$$\bar{K}_{\mathbf{k}}^{v v m f m'}(\infty) \sim e^{i\omega_{\text{pump}}\Delta\tau_{mm'}} \quad \text{(Floquet)}$$

$$\bar{K}_{\mathbf{k}}^{f v m f m'}(\infty) \sim e^{-i\omega_{\text{pump}}\Delta\tau_{mm'}} \quad \text{(Volkov)}$$

$$\bar{K}_{\mathbf{k}}^{c v m f m'}(\infty) \sim e^{i\omega_{\text{pump}}\Delta\tau_{mm'}} \quad \text{(VVCT).}$$

**Case 2.** For emission process ($l = 2$), Eq. (S19) is written as

$$\begin{aligned}\bar{K}_{\mathbf{k}}^{\mu v m f m'}(\infty) = \Bigg[&\frac{1}{i\omega_{\mathbf{k}}^{\mu v m} + \eta_m}\frac{e^{-(i\omega_{\mathbf{k}}^{\mu v m}+\eta_m)\Delta\tau_{mm'}}}{2i\omega_{\text{pump}} + \eta_m + \eta_{m'}} \\ &+ \frac{1}{i\omega_{\mathbf{k}}^{\mu v m} + \eta_m}\frac{e^{(2i\omega_{\text{pump}}-\omega_{\mathbf{k}}^{\mu v m}-\eta_{m'})\Delta\tau_{mm'}} - e^{-i(\omega_{\mathbf{k}}^{\mu v m}+\eta_m)\Delta\tau_{mm'}}}{2i\omega_{\text{pump}} + \eta_m - \eta_{m'}} \\ &- \frac{2\eta_m}{\omega_{\mathbf{k}}^{\mu v m\,2} + \eta_m^2}\frac{e^{(2i\omega_{\text{pump}}-\omega_{\mathbf{k}}^{\mu v m}-\eta_{m'})\Delta\tau_{mm'}}}{i(2\omega_{\text{pump}} - \omega_{\mu v m}^{\mathbf{k}}) - \eta_{m'}} \\ &- \frac{1}{i\omega_{\mathbf{k}}^{\mu v m} - \eta_m}\frac{e^{(2i\omega_{\text{pump}}-\omega_{\mathbf{k}}^{\mu v m}-\eta_{m'})\Delta\tau_{mm'}}}{2i\omega_{\text{pump}} - \eta_m - \eta_{m'}}\Bigg]\end{aligned} \quad \text{(S22)}$$

For emission processes of Floquet, Volkov, and VVCT, the following energy relations are obtained:

$$\text{Floquet: } \omega_{\mathbf{k}}^{\mu v m} = \omega_{\mathbf{k}}^{vv,\text{pump}} = \omega_{\text{pump}}$$

$$\text{Volkov: } \omega_{\mathbf{k}}^{\mu v m} = -\omega_{\mathbf{k}}^{\mu' \mu m'} + 2\omega_{\text{pump}} = \omega_{\text{pump}}$$

$$\text{VVCT: } \omega_{\mathbf{k}}^{\mu v m} = \omega_{\mathbf{k}}^{cv,\text{pump}} = \Delta E_{\mathbf{k}}^{cv} + \omega_{\text{pump}}.$$

Let us again approximate $\eta_{\text{pump}} \approx 0$ and $\frac{\omega_{\text{pump}}}{\eta_{XUV}} \approx 0$. Then, Eq. (S22) can be rewritten as

- Floquet & VVCT ($m = \text{pump}, m' = \text{probe}, \eta_m \to 0$):

$$K_{\mathbf{k}}^{\mu v m \mu' m'}(\infty) \approx -\frac{2ie^{-i\omega_{\mathbf{k}}^{\mu v m}\Delta\tau_{mm'}}}{\omega_{\mathbf{k}}^{\mu v m}\eta_{m'}}$$

- Volkov ($m = \text{probe}, m' = \text{pump}, \ \eta_{m'} \to 0$):

$$K_{\mathbf{k}}^{\mu v m \mu' m'}(\infty) \approx \frac{2ie^{\left(2i\omega_{\text{pump}}-\omega_{\mathbf{k}}^{\mu v m}\right)\Delta\tau_{mm'}}}{\left(2\omega_{\text{pump}}-\omega_{\mathbf{k}}^{\mu v m}\right)\eta_m}$$

Note that $\Delta\tau_{mm'} = -\tau_{\text{pump-probe}} \equiv -\tau$ for the Floquet and VVCT and $\Delta\tau_{mm'} = \tau$ for the Volkov. In both case 1 and case 2, it is found that $K_{\mathbf{k}}^{\mu v m \mu' m'}$ of Floquet and VVCT are proportional to $e^{i\omega_{\mathbf{k}}^{\mu v m}\tau}/\omega_{\mathbf{k}}^{\mu v m}$, and that of Volkov state is proportional to $\pm e^{\mp i\omega_{\text{pump}}\tau}/\omega_{\text{pump}}$. Remind that $\omega_{\mathbf{k}}^{\mu v m}$ for Floquet and VVCT are $\pm\omega_{\text{pump}}$ and $\Delta E_{\mathbf{k}}^{cv} \mp \omega_{\text{pump}}$, respectively. It indicates that $K_{\mathbf{k}}^{\mu v m \mu' m'}$ of every VVCT is proportional to $1/\left(\Delta E_{\mathbf{k}}^{cv} \mp \omega_{\text{pump}}\right)$, which is a positive value for both absorption and emission processes because $\omega_{\text{pump}} < \Delta E_{\mathbf{k}}^{cv}$. In contrast, $K_{\mathbf{k}}^{\mu v m \mu' m'}$ of Floquet and Volkov states are proportional to $\pm 1/\omega_{\text{pump}}$. Therefore, the interference of the absorption and emission process at RABBIT sidebands,

$$\left[K_{\mathbf{k}}^{\mu v m \mu' m'}(\infty)\right]_{\text{absorption}} + \left[K_{\mathbf{k}}^{\mu v m \mu' m'}(\infty)\right]_{\text{emission}}, \tag{S23}$$

gives $\pi$-phase shift for the VVCT compared to the Floquet and Volkov states. Note that the absolute phase of the Floquet, Volkov, and VVCT in a RABBIT spectrogram can be changed depending on the carrier envelope phase (CEP) of a whole APT. The $\pi$ phase shift indicates the relative phase between the Floquet-Volkov contribution and the VVCT contribution.

## S6. Emergence of topological features in RABBIT sidebands

In a RABBIT sideband, there is a complex quantum interference, including the Floquet, Volkov, VVCT contributions. According to *(3)*, matrix $\vec{\varepsilon}_{\text{pump}} \cdot \mathbf{M}_{\mathbf{k}}^{cv}$ in the VVCT state has an additional phase factor $e^{i\alpha_{\mathbf{k}}\Omega_{\mathbf{k}}}$ compared to Floquet and Volkov states where $\Omega_{\mathbf{k}}$ is the Berry curvature, and $\alpha_{\mathbf{k}}$ is a prefactor independent of the topology, as given in *(3)*. In addition, $\vec{\varepsilon}_{\text{pump}} \cdot \mathbf{M}_{\mathbf{k}}^{vv}$ and $\vec{\varepsilon}_{\text{pump}} \cdot \mathbf{M}_{\mathbf{k}}^{ff}$ are real and $\left(\vec{\varepsilon}_{\text{probe}} \cdot \mathbf{M}_{\mathbf{k}}^{fv}\right)/\left(\vec{\varepsilon}_{\text{probe}} \cdot \mathbf{M}_{\mathbf{k}}^{fc}\right)$ is almost real (i.e., real in the limit of the gapless graphene). Therefore, the Berry curvature and second order transition integral $K_{\mathbf{k}}^{\mu v m f m'}(\infty)$ plays a key role in the interference. Let us denote $\left\{\left|P_2^{\mathbf{k}}(\infty)\right\rangle,\ K_{\mathbf{k}}^{\mu v m f m'}(\infty)\right\}$ of Floquet, Volkov, and VVCT states be $\left\{\left|P_{\text{Floquet}}^{\mathbf{k}\pm}(\infty)\right\rangle,\ K_{\mathbf{k}}^{\text{Floquet}}\right\}$, $\left\{\left|P_{\text{Volkov}}^{\mathbf{k}\pm}(\infty)\right\rangle,\ K_{\mathbf{k}}^{\text{Volkov}}\right\}$, and $\left\{\left|P_{\text{VVCT}}^{\mathbf{k}\pm}(\infty)\right\rangle,\ K_{\mathbf{k}}^{\text{VVCT}}\right\}$, respectively, where $+(-)$ indicates the emission (absorption) processes. We take $m$ and $m'$ to be pump and an attosecond pulse of APT or opposite, respectively. As each attosecond pulse in Fig. S6 has a CEP shifted by $\frac{\pi}{2\omega_{\text{pump}}}$, we adopt $\omega_{\text{pump}}\Delta\tau_{mm'} \to \omega_{\text{pump}}\Delta\tau_{mm} + \frac{\pi}{2}$. According to the analysis in section S5, the total interference at a sideband is written as

$$\begin{aligned}\left|P_2^{\mathbf{k}}(\infty)\right\rangle &= \sum_{mm',s\in\{+,-\}}\left|P_{\text{Floquet}}^{\mathbf{k}s}(\infty)\right\rangle + \left|P_{\text{Volkov}}^{\mathbf{k}s}(\infty)\right\rangle + \left|P_{\text{VVCT}}^{\mathbf{k}s}(\infty)\right\rangle \\ &\approx \left(R_{\mathbf{k}}^{\text{FV}} + R_{\mathbf{k}}^{\text{VVCT}} e^{i\alpha_{\mathbf{k}}\Omega_{\mathbf{k}}}\right)\left|c_{f\mathbf{k}}\right\rangle\end{aligned} \tag{S24}$$

where $R_{\mathbf{k}}^{\text{FV}} \equiv -\frac{2}{\omega_{\text{pump}}}\left(D_{\text{Floquet}} + D_{\text{Volkov}}\right)\left(e^{i\omega_{\text{pump}}\Delta\tau_{mm'}} + e^{-i\omega_{\text{pump}}\Delta\tau_{mm'}}\right)$ and $R_{\mathbf{k}}^{\text{VVCT}} \equiv -2\bar{D}_{\text{VVCT}}\left[\frac{e^{i(\omega_{\text{pump}}-\Delta E_{\mathbf{k}}^{cv})\Delta\tau_{mm'}}}{\Delta E_{\mathbf{k}}^{cv}-\omega_{\text{pump}}} - \frac{e^{-(i\omega_{\text{pump}}+\Delta E_{\mathbf{k}}^{cv})\Delta\tau_{mm'}}}{\Delta E_{\mathbf{k}}^{cv}+\omega_{\text{pump}}}\right]$. Here, we define $D_{\text{Floquet}} = \left(A_0^{m'}\vec{\varepsilon}_{m'} \cdot \mathbf{M}_{\mathbf{k}}^{fv}\right)\left(A_0^{m}\vec{\varepsilon}_{m} \cdot \mathbf{M}_{\mathbf{k}}^{vv}\right)$, $D_{\text{Volkov}} = \left(A_0^{m'}\vec{\varepsilon}_{m'} \cdot \mathbf{M}_{\mathbf{k}}^{ff}\right)\left(A_0^{m}\vec{\varepsilon}_{m} \cdot \mathbf{M}_{\mathbf{k}}^{fv}\right)$, and $D_{\text{VVCT}} = \left(A_0^{m'}\vec{\varepsilon}_{m'} \cdot \mathbf{M}_{\mathbf{k}}^{fc}\right)\left(A_0^{m}\vec{\varepsilon}_{m} \cdot \mathbf{M}_{\mathbf{k}}^{cv}\right)$. Note that $D_{VVCT}$ is rewritten by $D_{VVCT} \equiv \bar{D}_{VVCT} e^{i\alpha_{\mathbf{k}}\Omega_{\mathbf{k}}}$ where $\bar{D}_{VVCT} = \left(A_0^{m'}\vec{\varepsilon}_{m'} \cdot \mathbf{M}_{\mathbf{k}}^{fc}\right)\left(A_0^{m}|\vec{\varepsilon}_{m} \cdot \mathbf{M}_{\mathbf{k}}^{cv}|\right)$ *(3)*. At the Floquet-Volkov dominant region, the second-order state is simplified as

$$\left|P_2^{\mathbf{k}}(\infty)\right\rangle = \sum_{mm',s=+,-}\left|P_{\text{Floquet}}^{\mathbf{k}s}(t)\right\rangle + \left|P_{\text{Volkov}}^{\mathbf{k}s}(t)\right\rangle + \left|P_{\text{VVCT}}^{\mathbf{k}s}(t)\right\rangle$$
$$\approx \frac{-2}{\omega_{\text{pump}}}\left(D_{\text{Floquet}} + D_{\text{Volkov}}\right)\left(e^{i\omega_{\text{pump}}\Delta\tau_{mm'}} + e^{-i\omega_{\text{pump}}\Delta\tau_{mm'}}\right)\left|c_{f\mathbf{k}}\right\rangle. \tag{S25}$$

and the RABBIT oscillation ($I_{\text{F}-\text{V}}^{\mathbf{k}} = \left|R_{\mathbf{k}}^{\text{FV}}\right|^2$) is obtained by

$$I_{\text{F}-\text{V}}^{\mathbf{k}} = \frac{8\left|D_{\text{Floquet}}+D_{\text{Volkov}}\right|^2}{\omega_{\text{pump}}^2}\left(1 + \cos 2\omega_{\text{pump}}\Delta\tau_{mm'}\right), \tag{S26}$$

which shows no time-shift. In contrast, if we only evaluate the VVCT contribution ($I_{\text{VVCT}}^{\mathbf{k}} = \left|R_{\mathbf{k}}^{\text{VVCT}}\right|^2$, we obtain

$$I_{\text{VVCT}}^{\mathbf{k}} = \frac{4|D_{\text{VVCT}}|^2}{\left(\Delta E_{\mathbf{k}}^{cv}-\omega_{\text{pump}}\right)^2} + \frac{4|D_{\text{VVCT}}|^2}{\left(\Delta E_{\mathbf{k}}^{cv}+\omega_{\text{pump}}\right)^2}$$
$$- \frac{8|D_{\text{VVCT}}|^2}{\left(\Delta E_{\mathbf{k}}^{cv}\right)^2-\omega_{\text{pump}}^2}\cos 2\omega_{\text{pump}}\Delta\tau_{mm'}. \tag{S27}$$

Note that equations S26 and S27 precisely explains each component RABBIT oscillation of Fig. 3 in the main text.

Noew, let us examine the interference between Floquet-Volkov and VVCT. As mentioned in the main text, $D_{\text{Volkov}}/D_{\text{Floquet}}$ is a real number and $\frac{D_{\text{VVCT}}}{D_{\text{Floquet}}}$ has an additional complex phase factor, i.e., $\frac{D_{\text{VVCT}}}{D_{\text{Floquet}}} = \left|\frac{D_{\text{VVCT}}}{D_{\text{Floquet}}}\right| e^{i\alpha_{\mathbf{k}}\Omega_{\mathbf{k}}}$. Therefore, the total interference at the RABBIT side band is written as

$$\left|R_{\mathbf{k}}^{\text{FV}} + R_{\mathbf{k}}^{\text{VVCT}}e^{i\alpha_{\mathbf{k}}\Omega_{\mathbf{k}}}\right|^2 = I_{\text{F}-\text{V}}^{\mathbf{k}} + I_{\text{VVCT}}^{\mathbf{k}} + 2\text{Re}\left[\left(R_{\mathbf{k}}^{\text{FV}}\right)^* R_{\mathbf{k}}^{\text{VVCT}}e^{i\alpha_{\mathbf{k}}\Omega_{\mathbf{k}}}\right]. \tag{S28}$$

After straightforward mathematical steps, the third term is evaluated as

$$2\text{Re}\left[\left(R_{\mathbf{k}}^{\text{FV}}\right)^* R_{\mathbf{k}}^{\text{VVCT}}e^{i\alpha_{\mathbf{k}}\Omega_{\mathbf{k}}}\right] = 4abc\cos\omega_{\text{pump}}\Delta\tau_{mm'}\cos\left(2\omega_{\text{pump}}\Delta\tau_{mm'} + \alpha_{\mathbf{k}}\Omega_{\mathbf{k}}\right)$$
$$4abd\cos\omega_{\text{pump}}\Delta\tau_{mm'}\cos(\alpha_{\mathbf{k}}\Omega_{\mathbf{k}}), \tag{S29}$$

where $a = -\frac{2}{\omega_{\text{pump}}}\left(D_{\text{Floquet}} + D_{\text{Volkov}}\right)$, $b = -2\bar{D}_{\text{VVCT}}$, $c = \frac{1}{\Delta E_{\mathbf{k}}^{cv} - \omega_{\text{pump}}}$, and $d = -\frac{1}{\Delta E_{\mathbf{k}}^{cv} + \omega_{\text{pump}}}$.

As $|c| \gg |d|$ at near resonant conditions, it is approximated as

$$2\text{Re}\left[\left(R_{\mathbf{k}}^{\text{FV}}\right)^{*} R_{\mathbf{k}}^{\text{VVCT}} e^{i\alpha_{\mathbf{k}}\Omega_{\mathbf{k}}}\right] \approx 4abc \cos \omega_{\text{pump}}\Delta\tau_{mm'} \cos\left(2\omega_{\text{pump}}\Delta\tau_{mm'} + \alpha_{\mathbf{k}}\Omega_{\mathbf{k}}\right). \quad \text{(S30)}$$

It proves that topological properties of RABBIT oscillation emerge due to the factor of $\cos\left[2\omega_{\text{pump}}\left(\Delta\tau_{mm'} + \frac{\alpha_{\mathbf{k}}\Omega_{\mathbf{k}}}{2\omega_{\text{pump}}}\right)\right]$.

## S7. Gauge invariance of the RABBIT spectroscopy

Let us consider an arbitrary Bloch gauge transformation, $|v_{\mathbf{k}}/c_{\mathbf{k}}\rangle \rightarrow |\bar{v}_{\mathbf{k}}/\bar{c}_{\mathbf{k}}\rangle \equiv e^{i\Phi_{v/c}(\mathbf{k})}|v_{\mathbf{k}}/c_{\mathbf{k}}\rangle$. It changes the second-order matrix element from $\left(\vec{\varepsilon}_{\text{probe}} \cdot \mathbf{M}_{\mathbf{k}}^{f\mu}\right)\left(\vec{\varepsilon}_{\text{pump}} \cdot \mathbf{M}_{\mathbf{k}}^{\mu v}\right)$ to $\left(\vec{\varepsilon}_{\text{probe}} \cdot \bar{\mathbf{M}}_{\mathbf{k}}^{f\mu}\right)\left(\vec{\varepsilon}_{\text{pump}} \cdot \bar{\mathbf{M}}_{\mathbf{k}}^{\mu v}\right)$. If $\mu \in \{v, c\}$ (i.e., Floquet and VVCT paths), the transformed dipole matrices are written as $\bar{\mathbf{M}}_{\mathbf{k}}^{\mu v} = \langle \bar{c}_{\mu\mathbf{k}} | \nabla_{\mathbf{k}} H_{\mathbf{k}} | \bar{c}_{v\mathbf{k}} \rangle = e^{-i\left(\Phi_{\mu}(\mathbf{k}) - \Phi_{v}(\mathbf{k})\right)} \mathbf{M}_{\mathbf{k}}^{\mu v}$ and $\bar{\mathbf{M}}_{\mathbf{k}}^{f\mu} = \langle c_{f\mathbf{k}} | \hat{\mathbf{k}} | \bar{c}_{v\mathbf{k}} \rangle = e^{i\Phi_{\mu}(\mathbf{k})} \mathbf{M}_{\mathbf{k}}^{f\mu}$, which result in $\left(\vec{\varepsilon}_{\text{probe}} \cdot \bar{\mathbf{M}}_{\mathbf{k}}^{f\mu}\right)\left(\vec{\varepsilon}_{\text{pump}} \cdot \bar{\mathbf{M}}_{\mathbf{k}}^{\mu v}\right) = e^{i\Phi_{v}(\mathbf{k})}\left(\vec{\varepsilon}_{\text{probe}} \cdot \mathbf{M}_{\mathbf{k}}^{f\mu}\right)\left(\vec{\varepsilon}_{\text{pump}} \cdot \mathbf{M}_{\mathbf{k}}^{\mu v}\right)$. If $\mu = f$ (i.e., Volkov state), then we similarly obtain $\left(\vec{\varepsilon}_{\text{pump}} \cdot \bar{\mathbf{M}}_{\mathbf{k}}^{ff}\right)\left(\vec{\varepsilon}_{\text{probe}} \cdot \bar{\mathbf{M}}_{\mathbf{k}}^{fv}\right) = e^{i\Phi_{v}(\mathbf{k})}\left(\vec{\varepsilon}_{\text{pump}} \cdot \mathbf{M}_{\mathbf{k}}^{ff}\right)\left(\vec{\varepsilon}_{\text{probe}} \cdot \mathbf{M}_{\mathbf{k}}^{fv}\right)$. That is, this gauge transform gives a global phase $e^{i\Phi_{v}(\mathbf{k})}$, which has no effect on the spectral intensity. Likewise, the gauge transformation of free electron final state also gives a global phase, which is cancelled out in the intensity calculation.

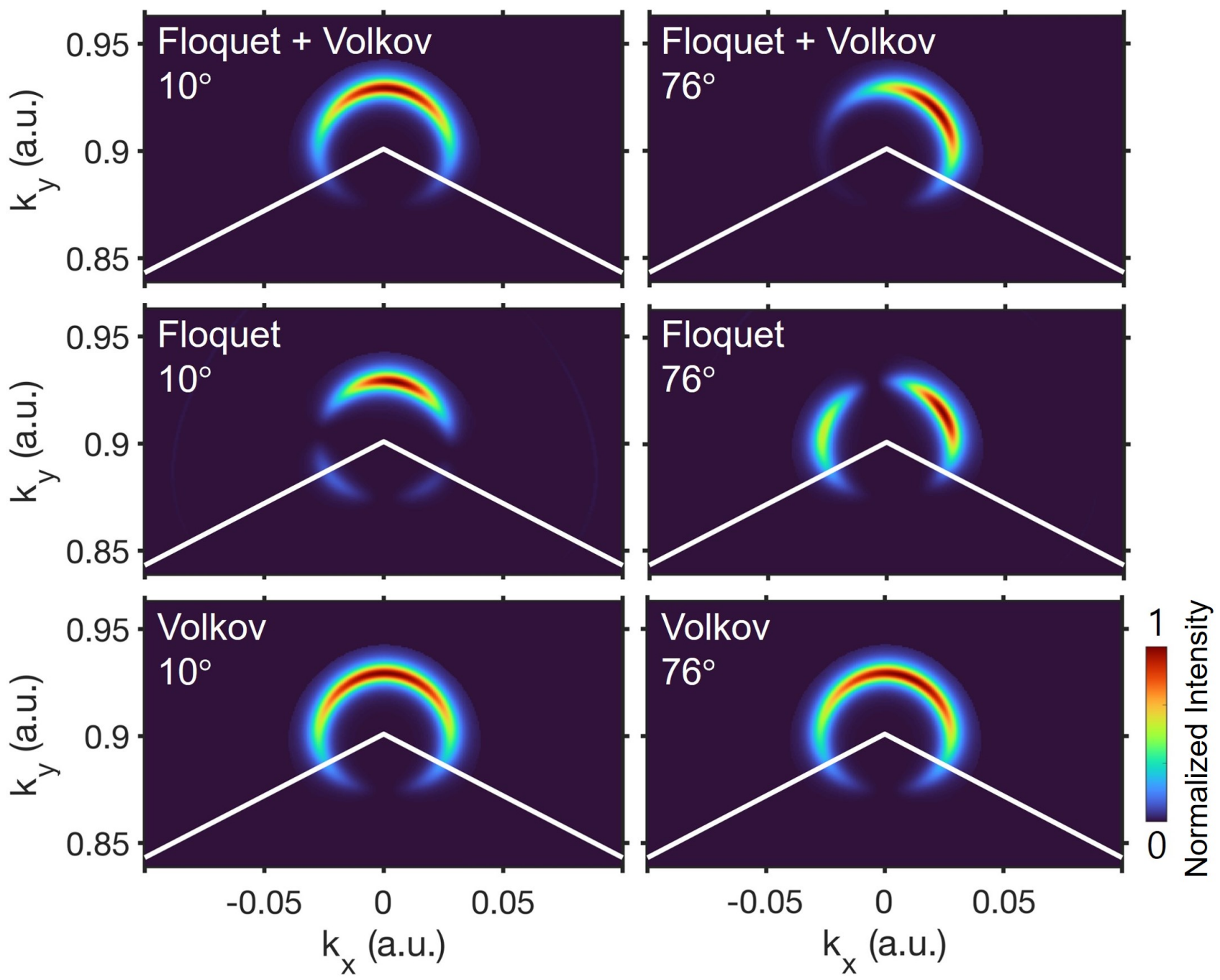

**Figure S1: Calculation of the recent Floquet-Volkov interference in doped graphene.** We follow experimental conditions of recent observations *(1,2)*. The intensity of each panel is normalized to its maximum value.

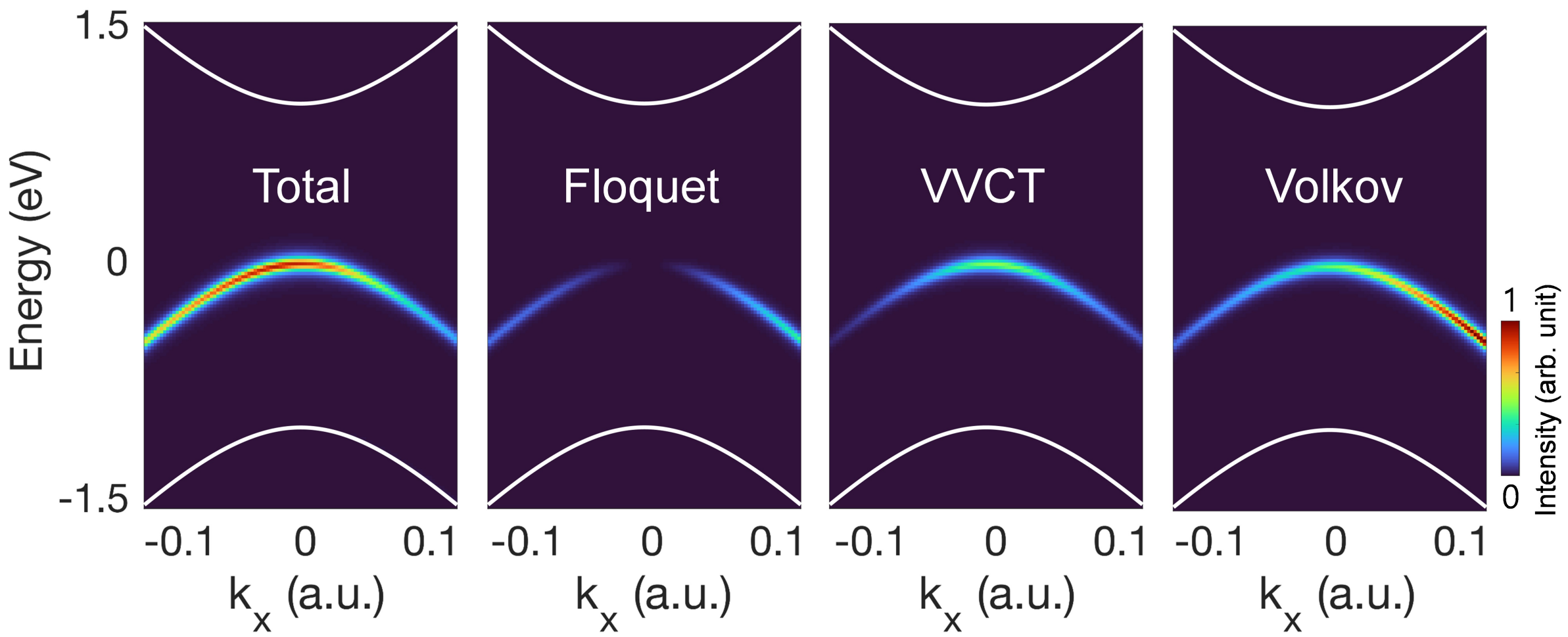

**Figure S2: Calculations of tr-ARPES for Floquet, VVCT, Volkov, and total spectra at $\boldsymbol{\tau_{\mathrm{pump-probe}}}$ = 0 fs.** It is performed by using second-order TDPT in a larger bandgap (= 2 eV) system. The pulse conditions are the same as those in Fig. 2(a) of the main text.

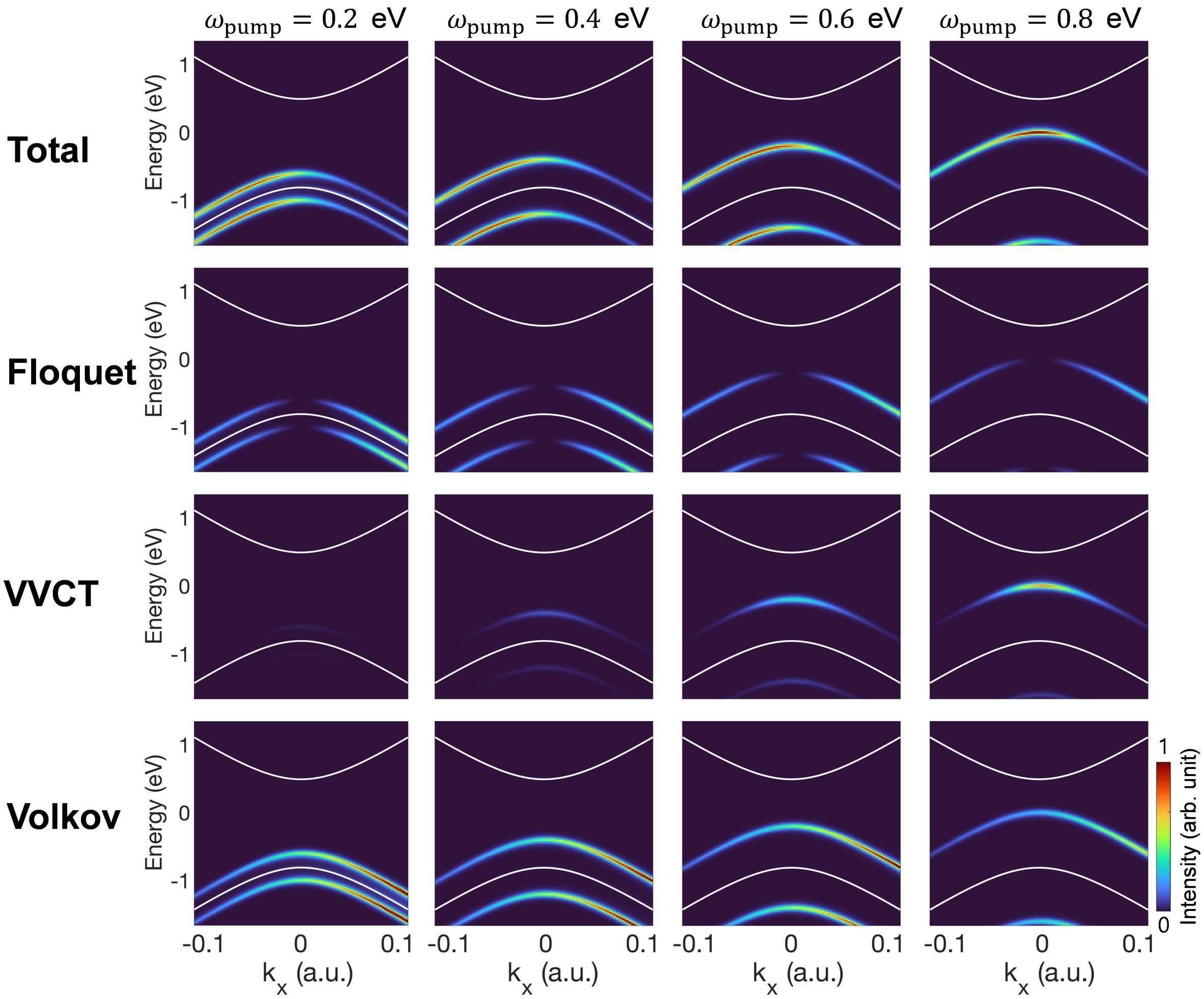

**Figure S3: Total, Floquet, VVCT, and Volkov features of tr-ARPES for various values of the pump frequency $\boldsymbol{\omega}_{\mathbf{pump}}$ at $\boldsymbol{\tau}_{\mathbf{pump-probe}} = \mathbf{0}$.** $E_g = 1.3$ eV is taken. The symmetry direction along which the calculation is done is same as in Fig.2A. The color range for each pump frequency is set from zero to the maximum intensity of the total spectra. The pulse conditions, except for the pump energy, are the same as those in Fig. 2A of the main text.

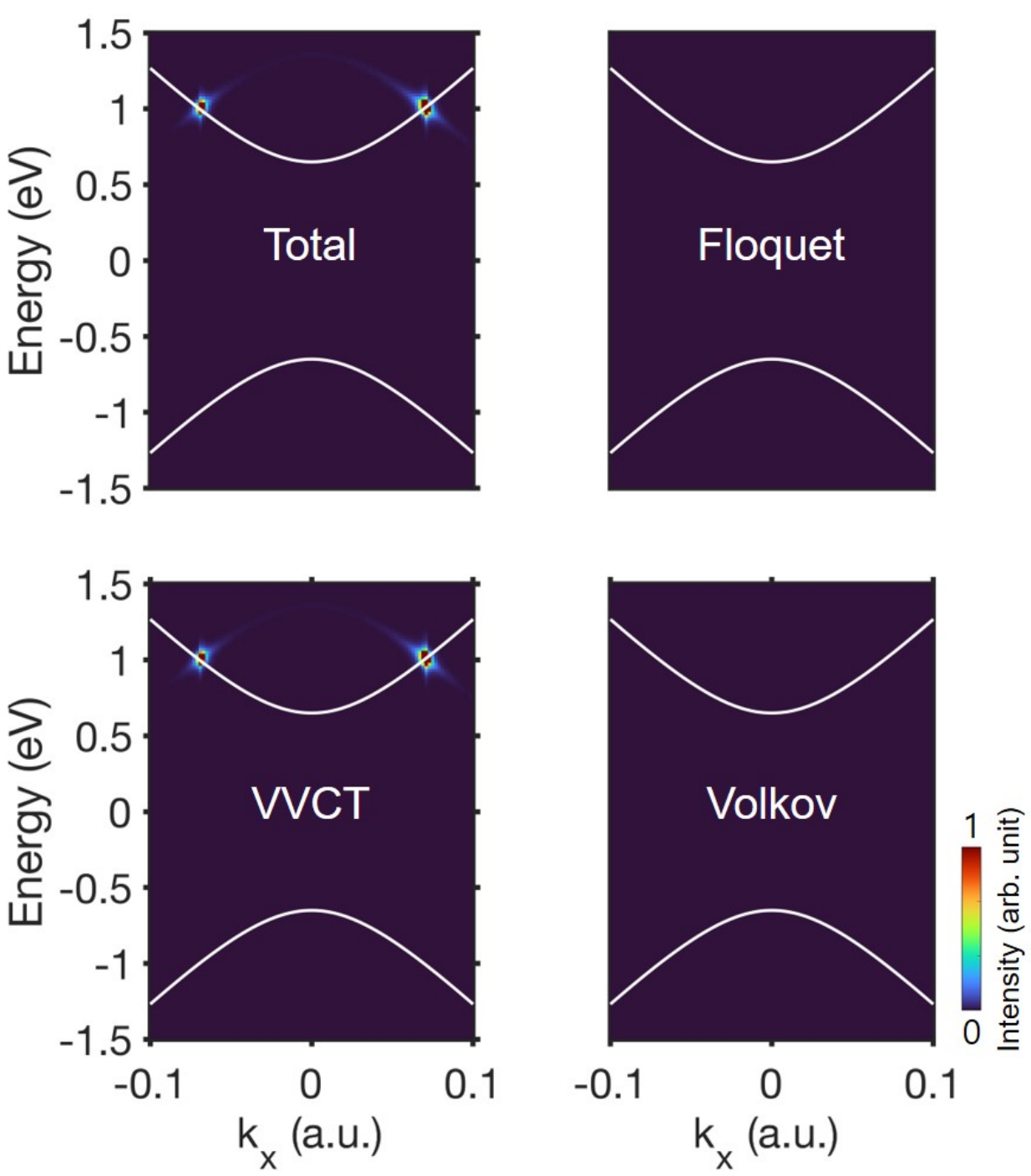

**Figure S4: Calculations of tr-ARPES for Floquet, VVCT, Volkov, and total spectra at $\boldsymbol{\tau_{\text{pump-probe}}} = \mathbf{0}$ fs.** It is performed by using second-order TDPT with an above-gap pump pulse, $\omega_{\text{pump}} = 2$ eV. The pulse conditions, except for the pump energy, are the same as those in Fig. 2A of the main text.

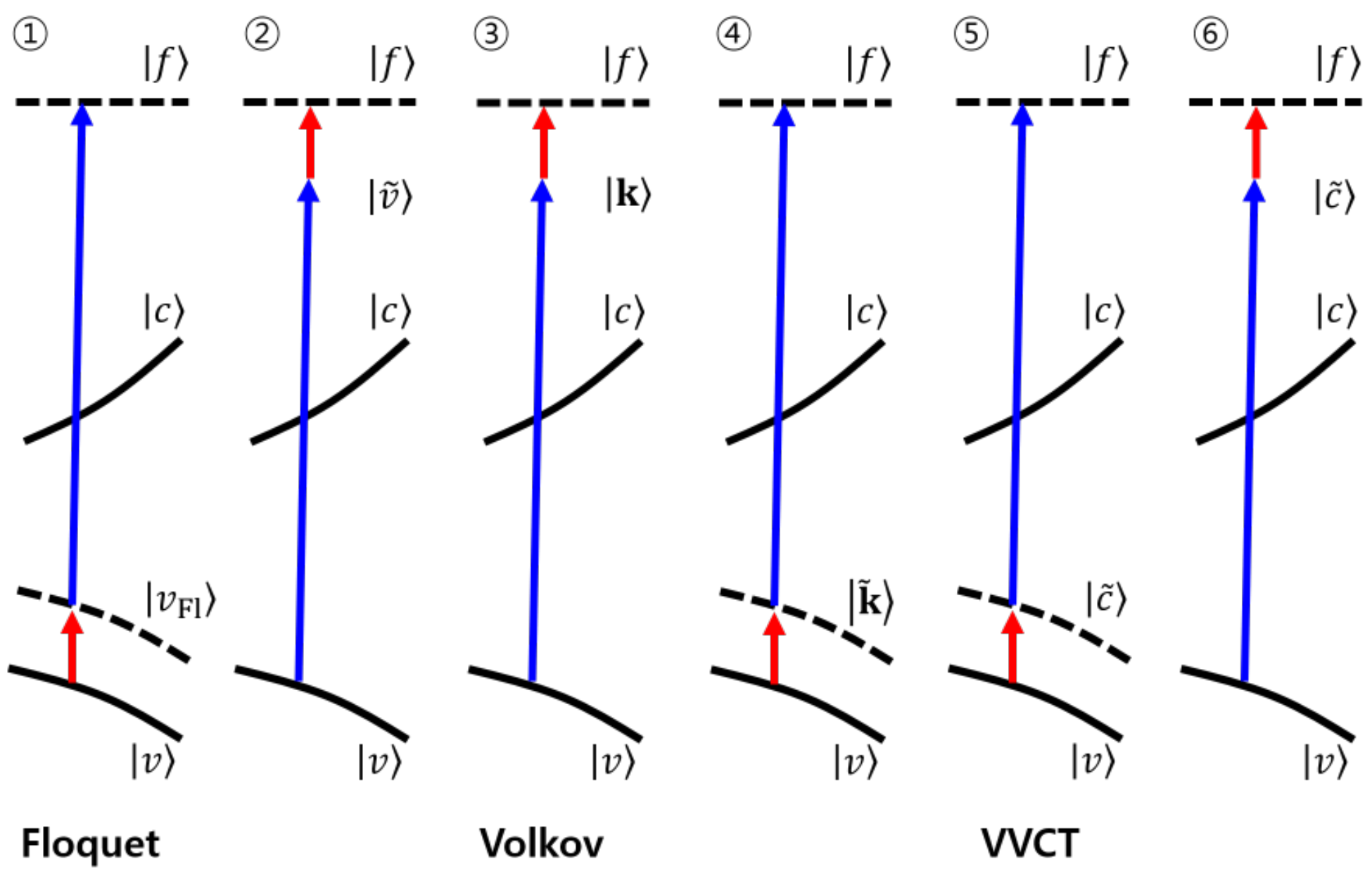

**Figure S5: Schematics of Floquet, Volkov, and VVCT states and the other second-order processes in the two-photon photoemission.** $|v\rangle$, $|c\rangle$, and $|\mathbf{k}\rangle$ are valence, conduction, and photoelectron states, respectively. Here, $|v_{\text{Fl}}\rangle$ is the first-order Floquet state of the valence state

and the tilde indicates the virtual transition. The red and blue arrows represent IR and XUV interactions, respectively.

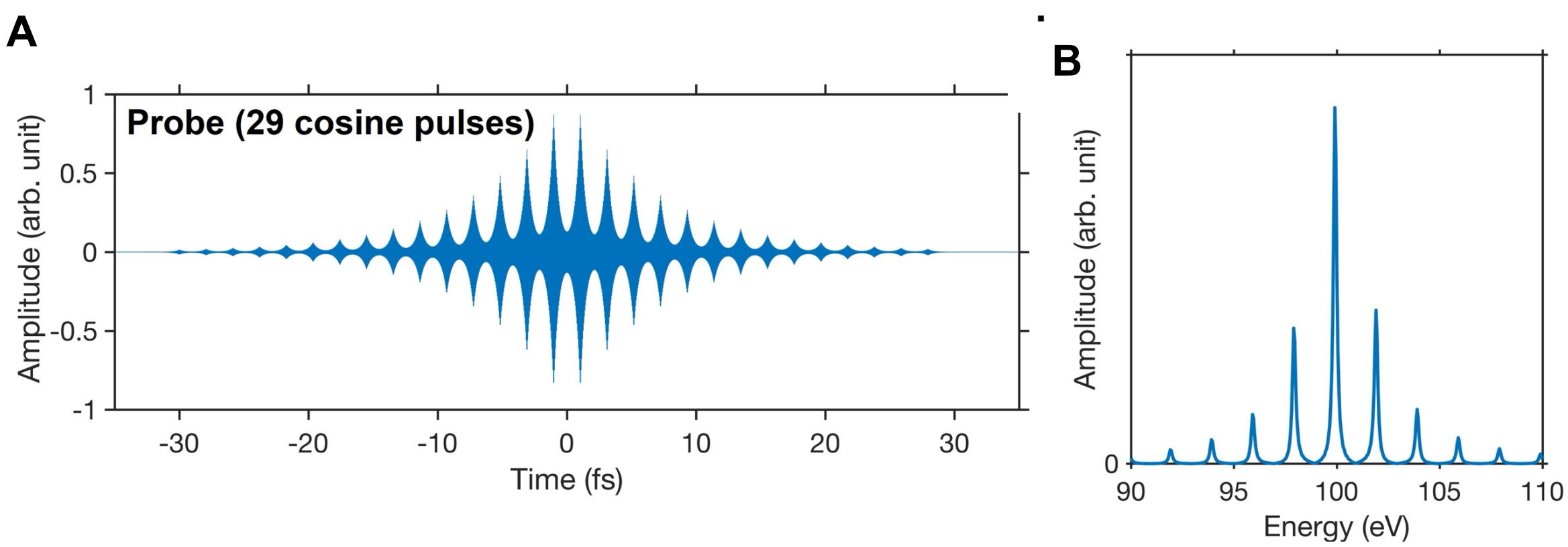

**Figure S6: The attosecond pulse train with multiple damped cosine pulses.** (**A**) Real-time plot of the attosecond pulse train with 29 damped cosine pulses mentioned in the main text. (**B**) The Fourier transformation of (**A**). The energy gap between the main peaks is 2 eV.

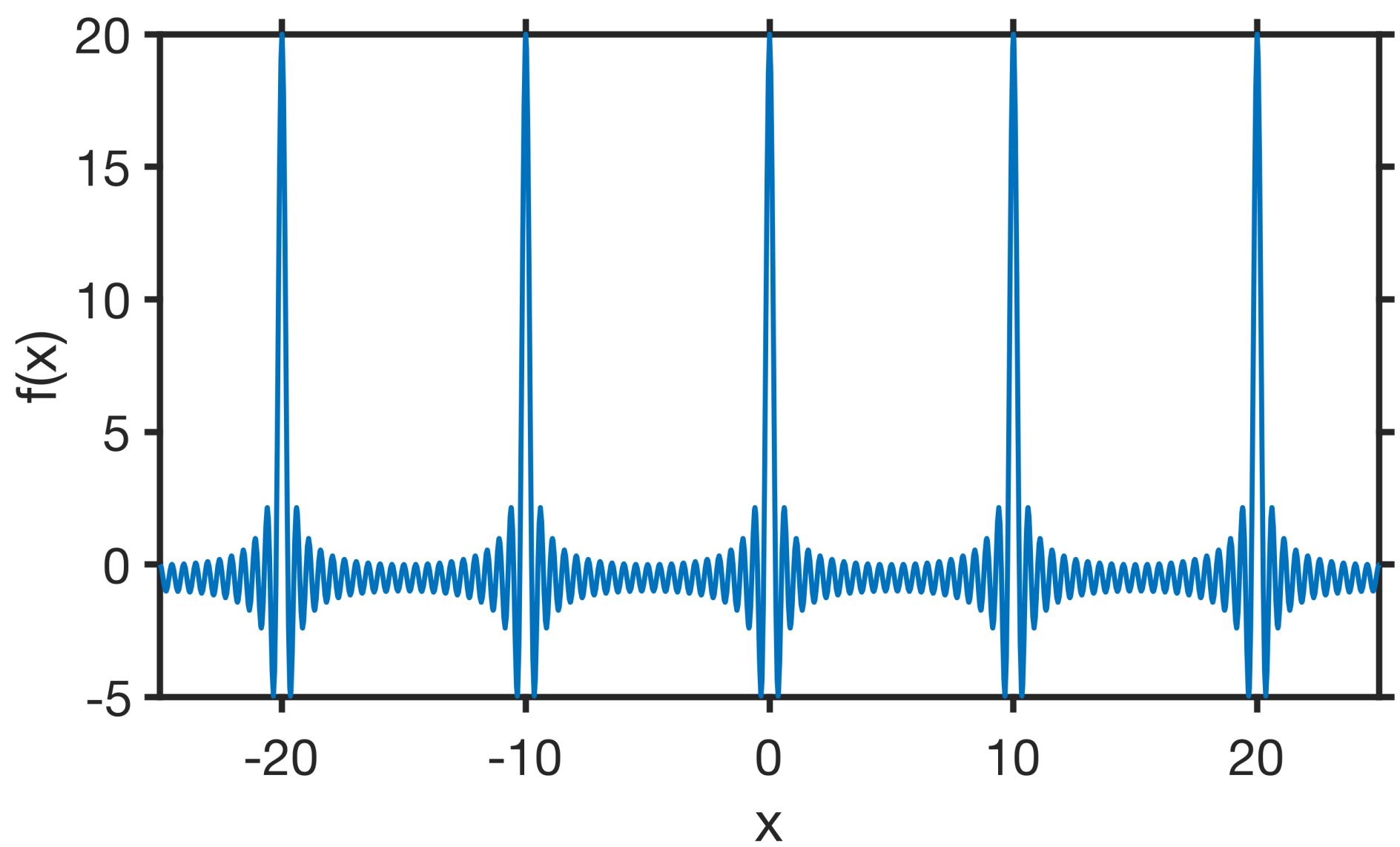

**Figure S7: Calculation of $\sum_{n=1} \cos\left(n\pi \frac{x}{x_0}\right)$ where $x_0 = 5$.** The summation is performed for $n = 1,2, \dots ,20$.

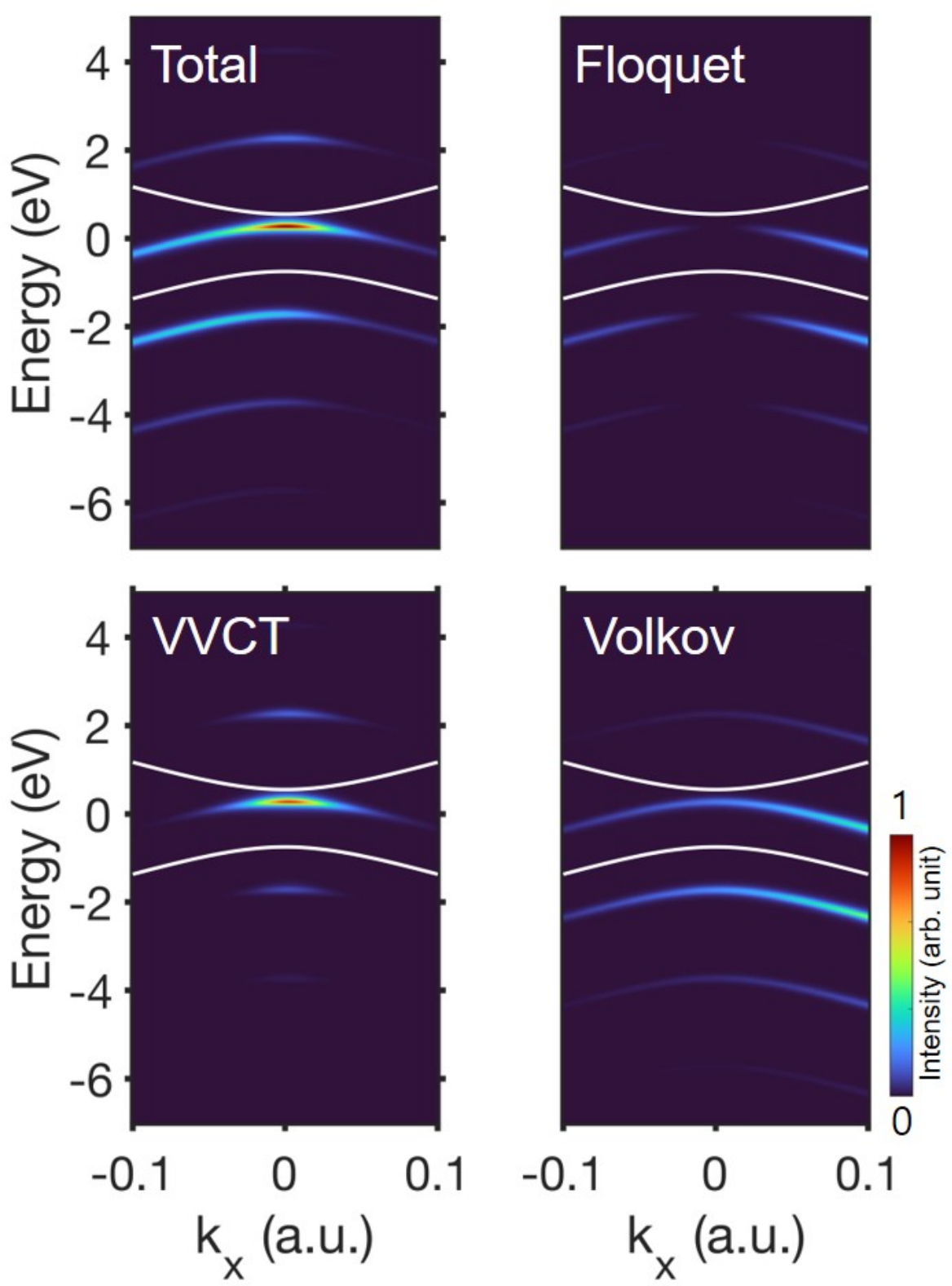

**Figure S8: The RABBIT sidebands in the momentum-energy grid of Fig. 3A along the k-line in the inset of Fig. 2.** All values are normalized by the maximum value of the total spectra.

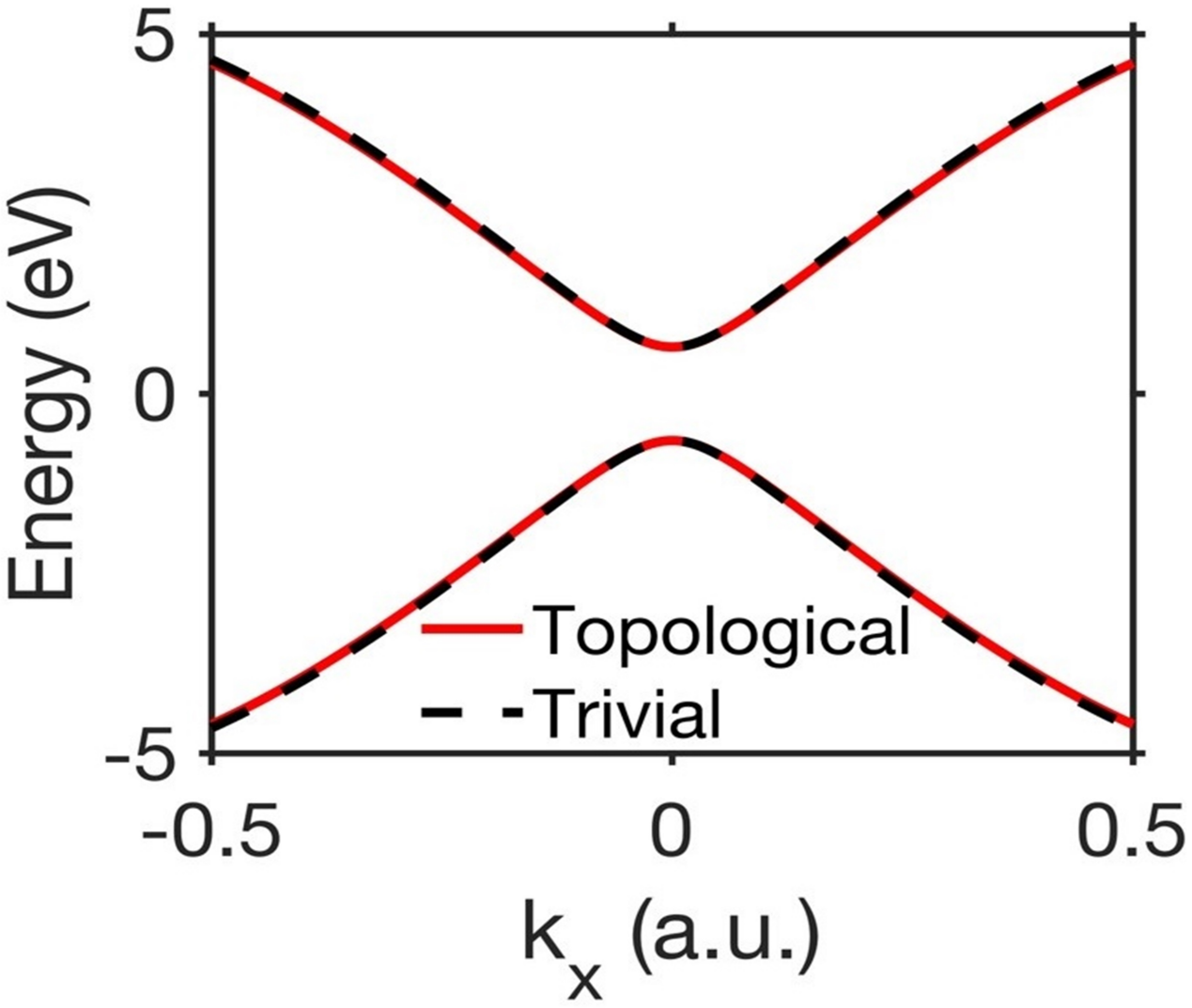

**Figure S9: Band structure of topologically trivial and nontrivial systems.** The detailed parameters are provided in section S1.

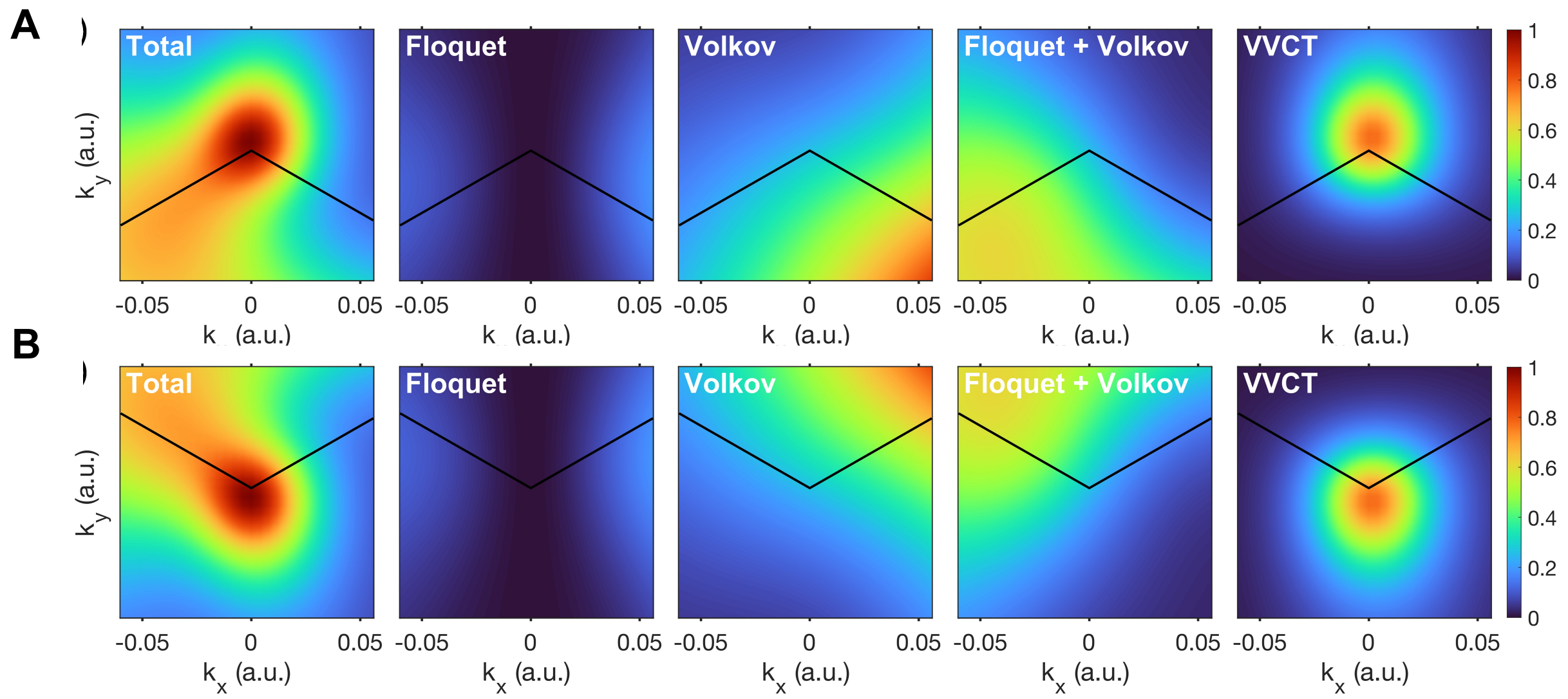

**Figure S10: Energy-integrated intensity of RABBIT sidebands of the topologically trivial system around K (a) and K′ (b) points.** Here, the intensity of total, Floquet, Volkov, Floquet + Volkov, and VVCT contribution in Fig. 4 are displayed. Almost the same results are obtained for the topologically nontrivial system.

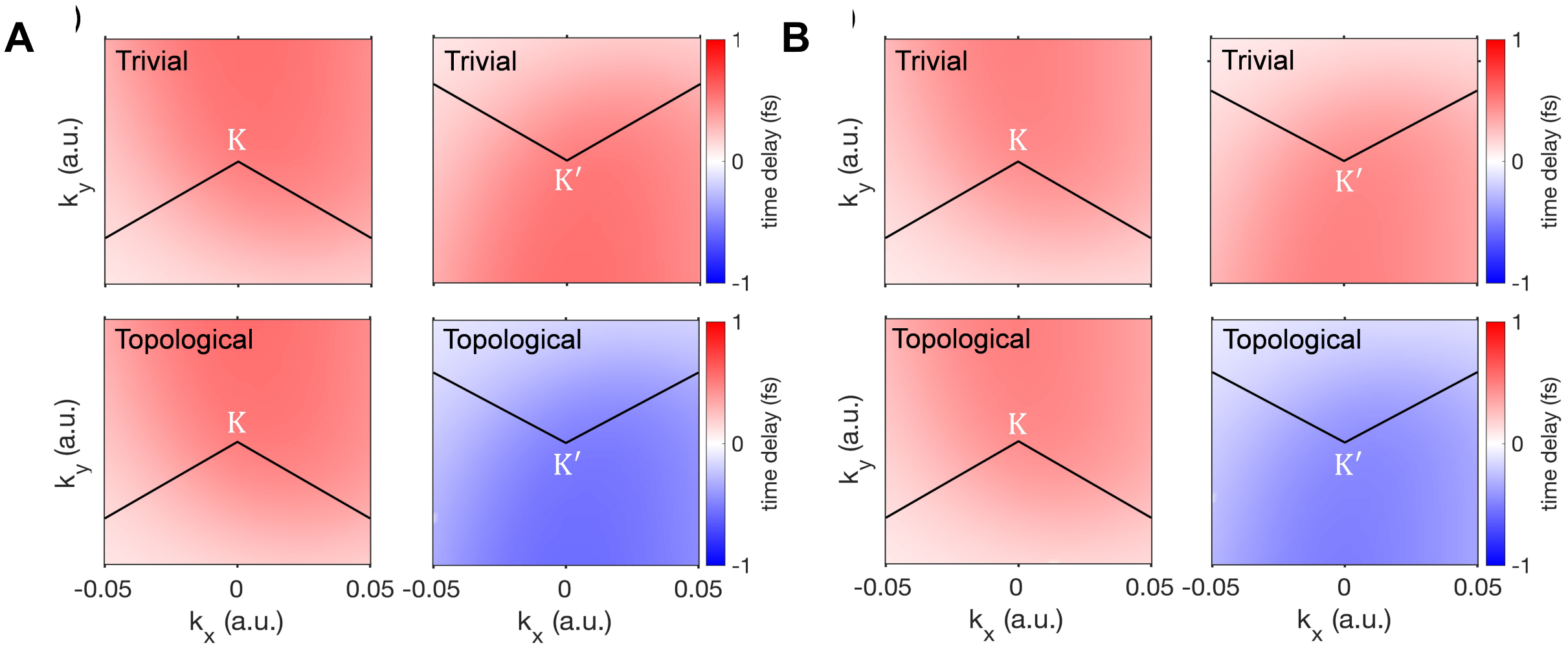

**Figure S11: Photoelectron emission delays around the points of trivial and nontrivial (topological) systems.** (**A**) The XUV (probe) energies of 200 eV and (**B**) 300 eV are adopted. The other pulse conditions are the same as those in Fig. 4 of the main text. Results are the same as Fig.4B-E of the main text, where the XUV (probe) energy of 100 eV is adopted.

**Caption for Movie S1: The total ARPES data with respect to the pump-probe delay for a single period of the pump pulse.** The parameters of the calculation are the same as those of Fig. 3 in the main text